\documentclass[twocolumn,showpacs,superscriptaddress,preprintnumbers,amsmath,amssymb,aps]{revtex4}
\usepackage{graphicx}
\usepackage{bm}

\newcommand{\ud}{\mathrm{d}}
\newcommand{\cref}[1]{(\ref{#1})}

\begin{document}
\title{A Triangular Tessellation Scheme for the Adsorption Free Energy at the Liquid-Liquid Interface: Towards Non-Convex Patterned Colloids}

\author{Joost de Graaf}%
 \email{j.degraaf1@uu.nl}
\author{Marjolein Dijkstra}%
\affiliation{%
Soft Condensed Matter, Debye Institute for Nanomaterials Science, Utrecht University, Princetonplein 5, 3584 CC Utrecht, The Netherlands
}%
\author{Ren\'e van Roij}%
\affiliation{%
Institute for Theoretical Physics, Utrecht
University, \\ Leuvenlaan 4, 3584 CE Utrecht, The Netherlands
}%

\date{\today}

\begin{abstract}
We introduce a new numerical technique, namely triangular tessellation, to calculate the free energy associated with the adsorption of a colloidal particle at a flat interface. The theory and numerical scheme presented here are sufficiently general to handle non-convex patchy colloids with arbitrary surface patterns characterized by a wetting angle, e.g., amphiphilicity. We ignore interfacial deformation due to capillary, electrostatic, or gravitational forces, but the method can be extended to take such effects into account. It is verified that the numerical method presented is accurate and sufficiently stable to be applied to more general situations than presented in this paper. The merits of the tessellation method prove to outweigh those of traditionally used semi-analytic approaches, especially when it comes to generality and applicability. 
\end{abstract}


\maketitle

\section{\label{sec:intro}Introduction}

Small particles at liquid-liquid interfaces are of scientific interest, but can also be exploited for industrial applications. Particles adsorbed at an interface have a multitude of applications, ranging from the formation of two-dimensional structures~\cite{pieran0,twoD}, which may be utilized in optical devices, to the stabilization of foams and pickering emulsions~\cite{stapick}. The range of sizes, shapes, and material properties with which colloids can be endowed, makes them the ideal constituents for self-assembled macroscopic structures. In addition, colloid tunability allows tailoring to specific systems, which gives tremendous advantages over atomic materials. A more fundamental impetus to the study of colloid adsorption, is based on gaining a better understanding in phase transitions and critical phenomena of two-dimensional fluids of nanoparticles at an interface.

Many theoretical investigations of colloids at an interface are based on studies into the behavior of a single particle at the interface. The stability of an adsorbed colloid and the manner in which it attaches to the interface gives insight into the way particles act at higher concentrations. The stability of colloids at an interface was already considered by Pieranski~\cite{pieran0}, who studied the adsorption free energy based on surface tension arguments. This ground breaking work was built upon to encompass effects, such as line tension~\cite{bresme,strange_th}, capillary rise~\cite{oettel1,scriven}, surface deformation due to gravity~\cite{neu1}, surface heterogeneities~\cite{JB0,microstr,neu2}, and electrostatic effects~\cite{dipping,oettel2}. The influence of particle shape on colloid adsorption has also been considered, for instance, ellipsoidal rods and platelets~\cite{oettel1,bresme}, and more complex shapes as well~\cite{strange}. Nevertheless, there are still many unanswered questions concerning the adsorption of a single particle at an interface.

To the best of our knowledge only one theoretical study has been undertaken into the effects of anisotropic particles adsorbed to the interface as a function of the particle's orientation~\cite{acicular}. Most studies have been limited to several mathematically convenient particle orientations, namely parallel or perpendicular to the interfacial normal~\cite{oettel0,oettel1,microstr,neu1,neu2,scriven,bresme,chan}. These orientations are also found in experimental systems~\cite{strange,strange_th,selfas,right,basa,vermant} and therefore the current theoretical descriptions give insight into the behavior of the particles. However, these insights are constrained to particles that remain in one of these orientations. Therefore, these theories cannot be used to analyze the mechanisms by which colloids end up in these orientations, or why these particular orientations are preferred over other orientations. 

Studying the free energy associated with the adsorption of an arbitrary shaped colloid with contact angle surface patterns is quite involved, especially when the colloid is allowed to have an arbitrary angle with the interface. We first examine homogeneous uni-axial convex colloids and formulate the adsorption free energy. Determining this adsorption free energy proves to be technically difficult for all but the most basic shapes. Therefore, we introduce a numerical technique, which we refer to as ``triangular tessellation'', to evaluate the adsorption free energy. The accuracy of this technique is verified by comparison with semi-analytic results for ellipsoids, cylinders, and spherocylinders. These semi-analytic results are derived by methods similar to those used in Ref.~\cite{acicular}. We improve some of these results. Furthermore, we extend the semi-analytic results of Ref.~\cite{acicular} to a wider class of particles. Finally, we formulate a theoretical description and present a numerical technique to handle non-convex colloids with surface patterns. 

In conclusion, we introduce a new numerical scheme to determine the adsorption free energy of non-convex particles with or without surface patterns at the liquid-liquid interface, which has many advantages, as regards, applicability, stability, and generality, over semi-analytic techniques used so far. More detailed studies based on this new technique will be presented elsewhere~\cite{unpub}.

\section{\label{sec:method}Method}

\subsection{\label{sub:theoretical}Theoretical Considerations}

\begin{figure}
\includegraphics[height=3.25in]{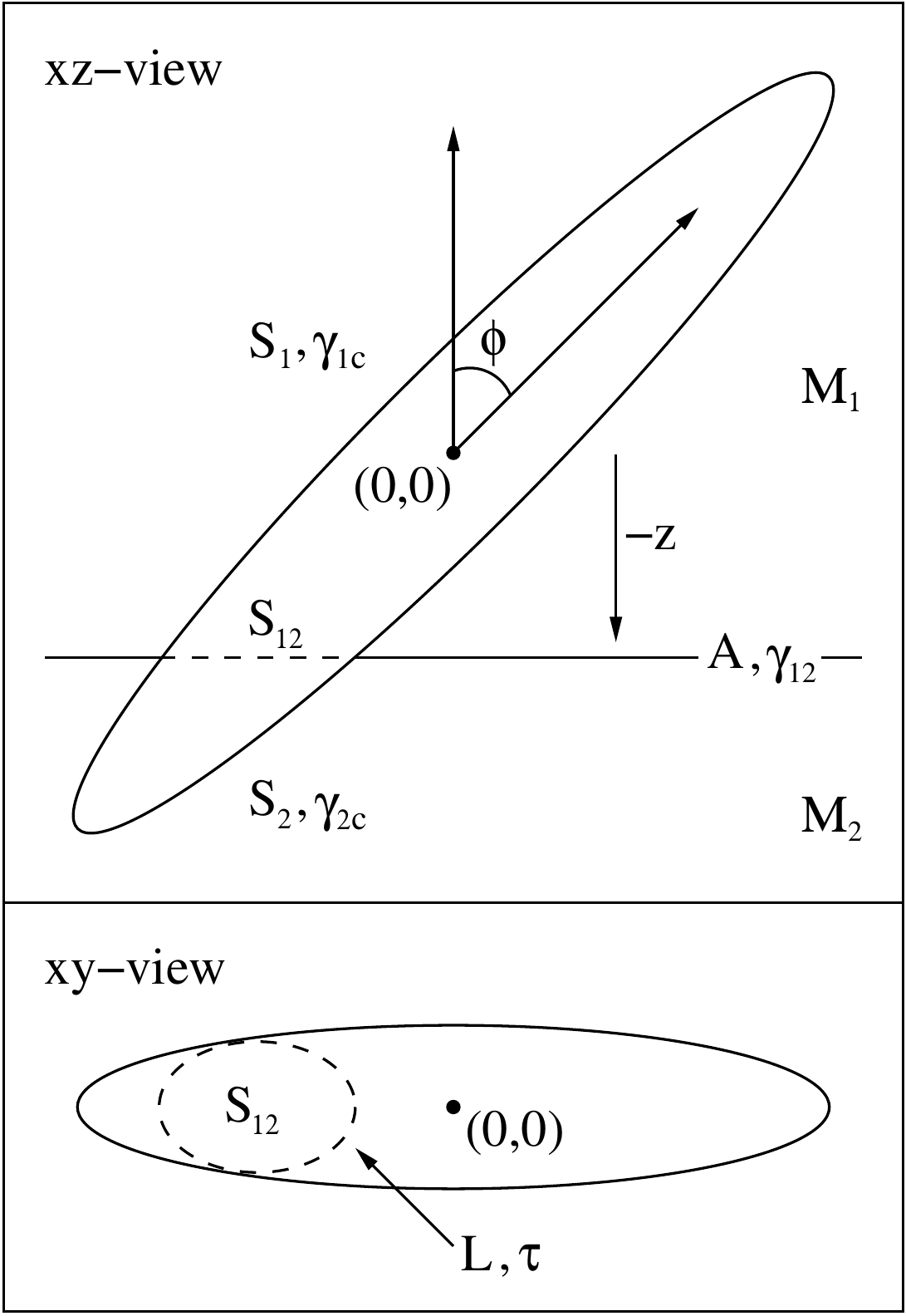}
\caption{\label{fig:config} A colloid (ellipsoid) adsorbed to the interface located at depth $-z$, measured from the center of the colloid $(0,0)$, dividing medium 1 ($M_{1}$) and medium 2 ($M_{2}$). The top frame shows an $xz$-plane cross section of the colloid and interface. The polar angle between the interfacial normal and the rotational symmetry axis of the colloid is denoted by $\phi$. The interface has total area $A$ with corresponding interfacial surface tension $\gamma_{12}$. The surface area of the colloid above the interface is denoted by $S_{1}$ with $\gamma_{1c}$ the $M_{1}$-colloid surface tension and the surface area of the colloid below the interface is denoted by $S_{2}$ with $\gamma_{2c}$ the $M_{2}$-colloid surface tension. The presence of the adsorbed colloid causes an area $S_{12}$ to be excluded from the interface, indicated by the dashed curve in the $xz$-view. In the bottom frame this excluded surface area is shown from an $xy$-view, i.e., the region enclosed by the dashed line. This dashed curve is the contact line, with length $L$ and corresponding line tension $\tau$. The outline of the colloid (solid curve) is also included.}
\end{figure}

We consider a planar oil-water interface separating two homogeneous half spaces of oil and water, and a solid uni-axial convex colloid adsorbed at this interface. We focus here on an oil-water interface, but we note that any liquid-liquid interface can be considered, and to some extent the theory is valid for liquid-gas interfaces as well. For simplicity, capillary effects due to the presence of the colloid at the interface are neglected. The coordinate frame is chosen such that the normal of the interface is along the $z$-axis. The position or \emph{depth} of the interface with respect to the center of the particle, at which the origin of the system is located, is denoted by depth $z$, which can be both positive and negative. For convenience we assume that the rotational symmetry axis of the particle is oriented in the $xz$-plane. The half space above the interface is called medium 1 ($M_{1}$) and the half space below the interface is called medium 2 ($M_{2}$), see Fig.~\ref{fig:config}. The angle between the colloid's rotational symmetry axis and interfacial normal is denoted by $\phi \in [0, \pi/2]$. Henceforth, $\phi$ is referred to as the \emph{polar angle}. 

There are four surface areas with corresponding surface tensions, which contribute to the adsorption free energy of the colloid: (i) the surface area of the colloid above the interface $S_{1}$, (ii) the surface area of the colloid below the interface $S_{2}$, (iii) the surface area excluded from the interface by the presence of the colloid $S_{12}$, and (iv) the total surface area of the interface (without adsorption) $A$. There is also a contribution from the contact line, of length $L$, where the three phases meet, i.e., $M_{1}$, $M_{2}$, and the colloid. Writing $S$ for the total surface area of the colloid, the following relations are obtained $S = S_{1} + S_{2}$ and $S_{1}, S_{2} \in [0,S]$.

These surface areas and the corresponding surface tensions, together with the contact line length and corresponding line tension give rise to an adsorption free energy. Such an adsorption free energy was first considered by Pieranski~\cite{pieran0} and later extended to accommodate line tension, e.g., Refs.~\cite{oettel0,bresme}. For a specific configuration, characterized by $z$ and $\phi$, this can be written as 
\begin{equation}
\label{eq:freeen}V(z,\phi)
 = \gamma_{12}(A - S_{12}) + \gamma_{1c}S_{1} + \gamma_{2c}S_{2} + \tau L,
\end{equation}
where $\gamma_{12}$ is the $M_{1}$-$M_{2}$ surface tension, $\gamma_{1c}$ is the $M_{1}$-colloid surface tension, $\gamma_{2c}$ is the $M_{2}$-colloid surface tension, and $\tau$ is the line tension. Note that we have dropped the $z$ and $\phi$ dependence of $S_{1}(z,\phi)$, $S_{2}(z,\phi)$, $S_{12}(z,\phi)$, and $L(z,\phi)$ to lighten the notation. The tensions (surface and line) are system parameters. 

The systems to which Eq.~\cref{eq:freeen} can be applied, have negligible capillary or gravitationally induced interfacial deformation, i.e., they have a flat interface and insignificant flotation force. The validity of such an assumption for colloidal systems can be studied using the Bond-number parameter, which gives the ratio between gravitational and surface tension induced effects on an adsorbing particle. Equation~\cref{eq:freeen} holds in the zero Bond-number limit
\begin{equation}
\label{eq:bond} \mathrm{Bo} = \frac{g \Delta \rho R^{2}}{\gamma_{12}} \rightarrow 0,
\end{equation}
where $\mathrm{Bo}$ is the Bond-number, $g$ is the gravitational acceleration, $\Delta \rho$ is the density difference between $M_{1}$ and $M_{2}$, and $R$ is the ``characteristic'' length scale associated with the particle. This result follows from conclusions in Refs.~\cite{neu1,chan}. The length scale $R$ is not entirely well-defined in the case of acicular (non-spherical) colloidal particles, but can be taken to be the mean radius of curvature. For colloids we estimate $R$ to be at most $10^{-5}$ m. The surface tension $\gamma_{12}$ should be at least $10^{-3}$ N~m$^{-1}$ for a liquid-liquid interface in the absence of surfactants, and the density difference can be at most of the order $10^{3}$ kg~m$^{-3}$ for physically reasonable systems. The Bond number is thus estimated to be $\mathrm{Bo} \ll 10^{-3}$. Hence it is safe to take the zero bond number limit~\cite{chan}, which implies insignificant interfacial deformation due to gravity. 

Capillary effects through immersion forces, anisotropy, and electrostatic interactions, cannot be so easily estimated. In Ref.~\cite{oettel0}, capillarity is discussed in the context of inter colloid interaction on the interface, however, the effects on a single particle's adsorption are not included. The theoretical validity of neglecting capillary deformation can be analyzed via de Young-Laplace equation, as done for a single orientation of ellipsoidal particles in Ref.~\cite{oettel1}. This analysis and the consequences for comparison with experimental systems, pertaining to the theoretical approach outlined in this paper, is left for future study~\cite{unpub}.

It is customary to define the adsorption free energy with respect to a reference point. The shifted adsorption free energy $F(z,\phi)$ is introduced by modifying $V(z,\phi)$ in such a way that it is zero when the colloid is completely immersed in $M_{1}$, i.e., by subtracting $\gamma_{12}A + \gamma_{1c}S$. This yields
\begin{equation}
\label{eq:freeenergy}F(z,\phi) = (\gamma_{1c} - \gamma_{2c})(S_{1} - S) - \gamma_{12}S_{12} + \tau L.
\end{equation}
Often, the term ``shifted'' is ignored when referring to $F(z,\phi)$. Only systems with $M_{1} \ne M_{2}$ are considered, hence $\gamma_{12} \ne 0$. Therefore it is possible to write $\gamma_{12} \cos\theta = \gamma_{1c} - \gamma_{2c}$, where the \emph{contact angle} $\theta$ is introduced via Young's equation~\cite{young}. Using this definition Eq.~\cref{eq:freeenergy} is rewritten to
\begin{equation}
\label{eq:free}F(z,\phi) = \gamma_{12}[(S_{1} - S)\cos\theta - S_{12}] + \tau L.
\end{equation}
Note that the contact angle $\theta$ is a quantity, which depends on the physical properties of the three components present at the interface, whereas the polar angle $\phi$ is a degree of freedom.

Dividing the adsorption free energy by $\gamma_{12}S$ ($\gamma_{12} \ne 0$) and writing $z = z^{*} \sqrt{a^{2}+2b^{2}}$, with $a$ the rotational symmetry semi-axis, $b$ the perpendicular semi-axis, and $m \equiv a/b$ the \emph{aspect ratio}, the following dimensionless adsorption free energy is derived
\begin{equation}
\label{eq:pierpot} f(z^{*},\phi) = \frac{F(z,\phi)}{\gamma_{12}S}  = \cos\theta(r_{1}-1) - r_{12} + \tau^{*} l,
\end{equation}
where $r_{1} \equiv S_{1}/S$ and $r_{12} \equiv S_{12}/S$ are surface area ratios,
\begin{equation}
\label{eq:taustar} \tau^{*} \equiv \frac{\tau}{\gamma_{12} \sqrt{S}},
\end{equation}
is the dimensionless line tension, and $l \equiv L/\sqrt{S}$ is a dimensionless contact line length. The value $\sqrt{a^{2}+2b^{2}}$ is the length of the semi-diagonal of a rectangular beam with sides $2a\times2b\times2b$. Two inequalities, namely $0 \le r_{1}(z^{*},\phi) \le 1$ and $0 \le r_{12}(z^{*},\phi) < 1$, hold for any value of $z^{*}$ and $\phi$ values. Note that $f(z^{*},\phi)$ implicitly depends on $\cos\theta$, $m$, and the shape of the particle.

The dimensionless adsorption free energy, Eq.~\cref{eq:pierpot}, is scale invariant, i.e., it is independent of the size of the colloid. For $\tau^{*} = 0$ our results for Eq.~\cref{eq:pierpot} hold for any size colloid. For $\tau^{*} \ne 0$ the results for Eq.~\cref{eq:pierpot} can be translated back to any colloid size, under the condition that for each size the line tension should be scaled. In this paper we focus on Eq.~\cref{eq:pierpot} rather than Eq.~\cref{eq:free}, since our goal is to develop a general method to describe the adsorption of arbitrary colloid shapes at an interface. Therefore, semi-analytic and fully numerical results are compared for the dimensionless adsorption free energy. 

Before we present our numerical technique, we first introduce some quantities which prove to be useful in describing the results. The location of the adsorption free energy minimum in Eq.~\cref{eq:pierpot} is denoted by $(z_{\mathrm{ad}}^{*},\phi_{\mathrm{ad}})$, which is referred to as the \emph{adsorption orientation}. The corresponding adsorption free energy reads $f_{\mathrm{ad}} \equiv f(z_{\mathrm{ad}}^{*},\phi_{\mathrm{ad}})$. Note that there may be multiple minima, in which case there can be meta-stable adsorption orientations. Several minima are labeled with a subscript $i = 1, 2, \dots$, where the deepest minimum has the lowest index. When there are two or more minima with equal adsorption free energy, we label them arbitrarily. 

For a given $\phi$, $z^{*}_{\mathrm{det}}(\phi)$ is defined as the positive value of $z^{*}$ for which the interface just touches the top of the particle. The colloid is detached when $z^{*} < -z^{*}_{\mathrm{det}}(\phi)$ or $z^{*} > z^{*}_{\mathrm{det}}(\phi)$. The quantity $z^{*}_{\min}(\phi)$ is defined as the value for which $f(z^{*},\phi)$ assumes its minimum as a function of $z^{*}$ for a given $\phi$. The corresponding adsorption free energy is denoted by $f_{\min}(\phi) \equiv f(z^{*}_{\min}(\phi),\phi)$. Note that it is \emph{a priori} not excluded that for a given $\phi$ the equi-$\phi$-curve has two or more (meta-stable) minima. This can correspond to multiple $z^{*}_{\min}(\phi)$ curves in the free energy landscape, running ``side-by-side'' in the $\phi$ direction. Often we will write $z_{\mathrm{det}}^{*}$ and $z_{\min}^{*}$ for $z_{\mathrm{det}}^{*}(\phi)$ and $z_{\min}^{*}(\phi)$ respectively, taking the polar angle dependence to be implicit.

In summary, we have described our theoretical model for a colloidal particle at a planar interface based on free energy arguments in the zero Bond number limit. This model can be straightforwardly generalized to encompass more complex colloidal properties such as surface patterns, see Appendix A.  Including the effects of gravity and capillary interfacial deformation is, however, substantially more involved and therefore not pursued here.

\subsection{\label{sub:numerical}Numerical Approximation Scheme}

\begin{figure}
\includegraphics[width=3.0in]{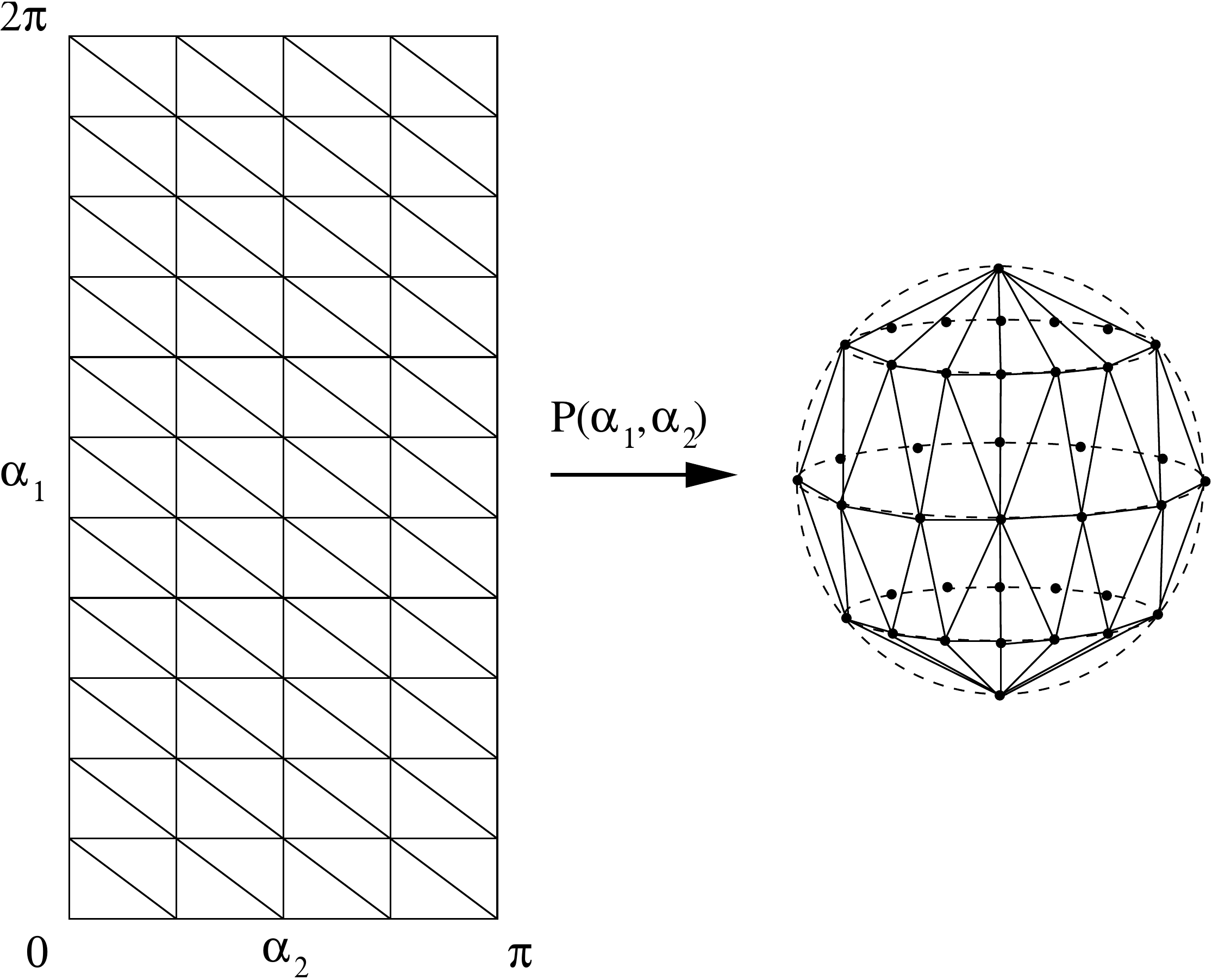}
\caption{\label{fig:para} Example of a parametrization $P(\alpha_{1},\alpha_{2})$ mapping the vertices of triangles in the strip $(\alpha_{1},\alpha_{2}) \in [0,2\pi]\times[0,\pi]$ to points on a sphere. In between these points ``mapped'' triangles are formed corresponding to the original triangles on the strip. The points at the poles are degenerate, all vertices with $\alpha_{2}=0$ and $\alpha_{2}=\pi$ coincide at the corresponding pole. Only the front half of the sphere
 is tessellated for clarity.}
\end{figure}

Determining the dependence of $S_{1}$, $S_{2}$, $S_{12}$, and $L$ on $z^{*}$ and $\phi$ is highly non-trivial in general, and deriving analytic expressions is unpractical, if not impossible, for all but the simplest particle shapes and orientations, see Appendix B. To analyze colloids adsorbed at an interface the following numerical technique is employed. The surface of the colloid is bijectively parametrized by two angles, namely $\alpha_{1} \in [0,2\pi]$ (azimuthal) and $\alpha_{2} \in [0,\pi]$ (polar). A parametrization can for instance take the form
\begin{equation}
\label{eq:ang_para} P(\alpha_{1},\alpha_{2}) = \left( \begin{array}{l} r(\alpha_{1},\alpha_{2}) \cos \alpha_{1} \sin \alpha_{2} \\ r(\alpha_{1},\alpha_{2}) \sin \alpha_{1} \sin \alpha_{2} \\ r(\alpha_{1},\alpha_{2}) \cos \alpha_{2} \\ \end{array} \right) ,
\end{equation}
where $r(\alpha_{1},\alpha_{2})$ is some radial function, but many other forms are imaginable. The strip $[0,2\pi]\times[0,\pi]$ is divided into triangles, the vertices of which are mapped onto the surface of the particle by means of the parametrization $P(\alpha_{1},\alpha_{2})$, see Fig.~\ref{fig:para}. A \emph{mapped triangle} is formed between the vertices of a corresponding triangle in the strip after $P$ has acted on them. From now on, the object, on which the strip's triangle mesh is mapped, is referred to as being \emph{tessellated} with triangles. The above method of modeling a 2D or 3D object by triangles (more generally polygons) is well-known in computer science and has been successfully applied to various surface tension problems in physics~\cite{brakke0,brakke1,brakke2}.

The surface area of the colloid is now approximated by summing the surface areas of the mapped triangles.  Suppose that the vertices of a mapped triangle are given by $\mathbf{x}$, $\mathbf{y}$, and $\mathbf{z}$, then its surface area is given by a simple cross-product $\vert (\mathbf{z - x})\times (\mathbf{y - x}) \vert/2$. This procedure
 can yield ``in principle'' arbitrary precision by sufficiently refining the triangular mesh. It should be noted
 that depending on the parametrization some triangles have a vanishingly small or zero contribution to the surface area. For example, in the case of a sphere several vertices coincide resulting in degenerate triangles at the poles, see Fig.~\ref{fig:para}. Note that this mapping is not bijective, but only on a set of which the image has measure zero. 

The method described above can be amended in the following manner to enable calculation of $S_{1}$, $S_{2}$, $S_{12}$ and $L$. Suppose that the tessellated object is intersected by a plane, then some of the triangles which compose the object lie above it and others below it. Let $\Delta_{\uparrow}$ denote the set of triangles which lie strictly above, $\Delta_{\downarrow}$ the set of triangles which lie strictly below, and $\Delta_{p}$ the set of triangles which intersect the plane or touch it. The surface of the colloid is approximated by $\tilde{S} = \sum_{\forall i}\Delta_{\uparrow,i} + \sum_{\forall j}\Delta_{\downarrow,j} + \sum_{\forall k}\Delta_{p,k}$, where the tilde indicates that this is an approximation, $i$, $j$, and $k$ are indices, and the notation for an element in a set doubles as the notation for that triangles surface area.

\begin{figure}
\includegraphics[width=2.95in]{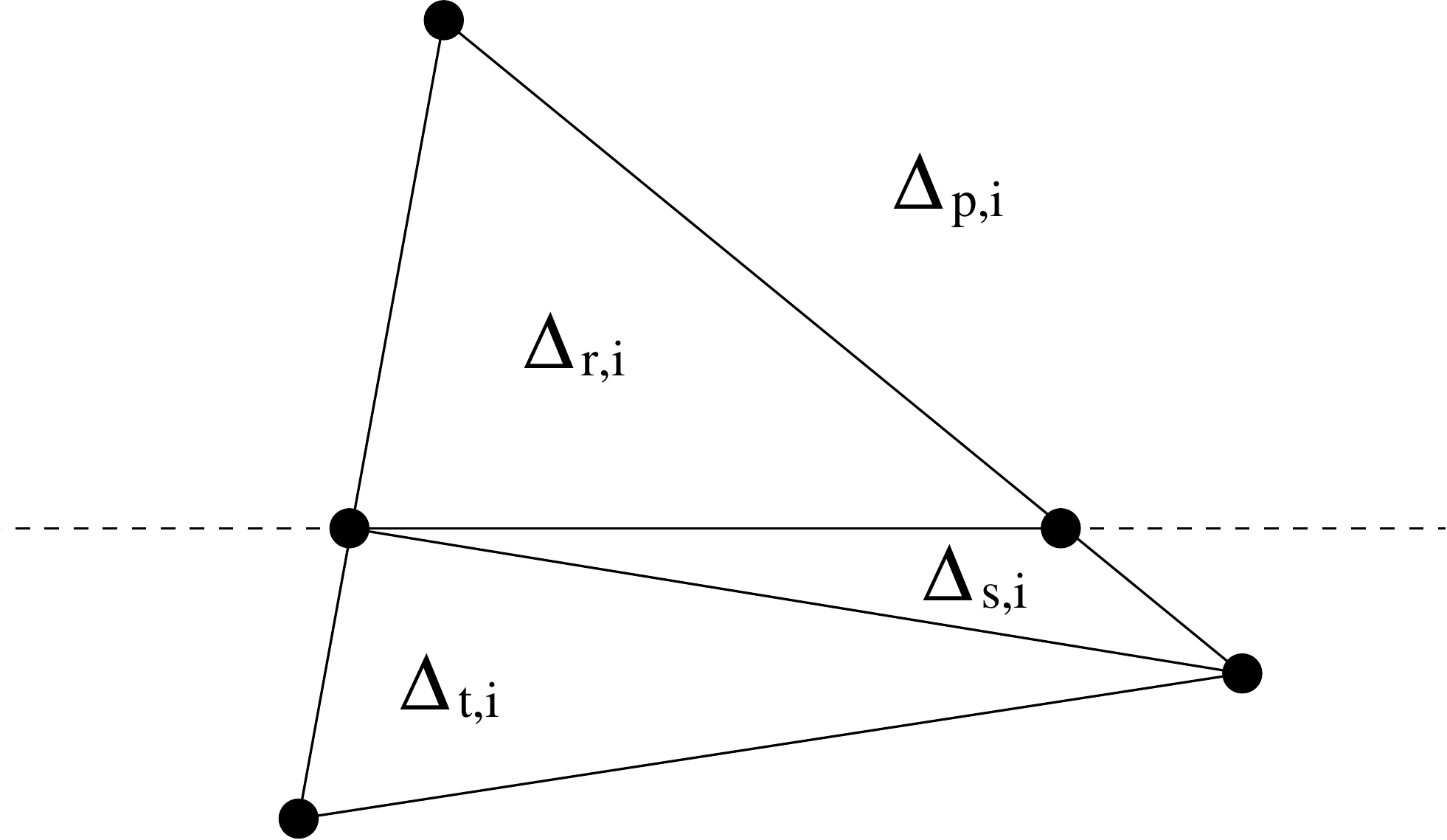}
\caption{\label{fig:cut} Example of a triangle $\Delta_{p,i}$ which intersects the interface (dashed line). Such a triangle can be cut into three pieces, $\Delta_{r,i}$, $\Delta_{s,i}$, and $\Delta_{t,i}$ as indicated above. In this case, the first piece lies above the interface and the second and third piece below it. The two new vertices lie on the interface.}
\end{figure}

Each intersected triangle $\Delta_{p,i}$ is divided into three sub-triangles $\Delta_{r,i}$, $\Delta_{s,i}$, and $\Delta_{t,i}$ in the manner indicated in Fig.~\ref{fig:cut}. Two of these lie on one side of the plane, and one on the other. Applying this technique to all triangles in $\Delta_{p}$ a set of partitioned triangles $\tilde{\Delta}_{p}$ is obtained, of which the members only have some vertices in common with the plane and do not intersect it. Let $\tilde{\Delta}_{p,\uparrow}$ and $\tilde{\Delta}_{p,\downarrow}$ be the sets of triangles in $\tilde{\Delta}_{p}$ which lie above and below the interface respectively, and let $\tilde{\Delta}_{\uparrow} = \Delta_{\uparrow}\cup\tilde{\Delta}_{p,\uparrow}$ and $\tilde{\Delta}_{\downarrow} = \Delta_{\downarrow}\cup\tilde{\Delta}_{p,\downarrow}$. Using these sets of triangles the surface areas $S_{1}$ and $S_{2}$ are approximated in the following way
\begin{eqnarray}
\label{eq:S1} \tilde{S}_{1} & = & \sum_{\forall i} \tilde{\Delta}_{\uparrow,i}, \\
\label{eq:S2} \tilde{S}_{2} & = & \sum_{\forall i} \tilde{\Delta}_{\downarrow,i}.
\end{eqnarray}
Note that by virtue of this technique the equality $\tilde{S} = \tilde{S}_{1} + \tilde{S}_{2}$ still holds, which can be used as a consistency check.

\begin{figure}
\includegraphics[width=2.4in]{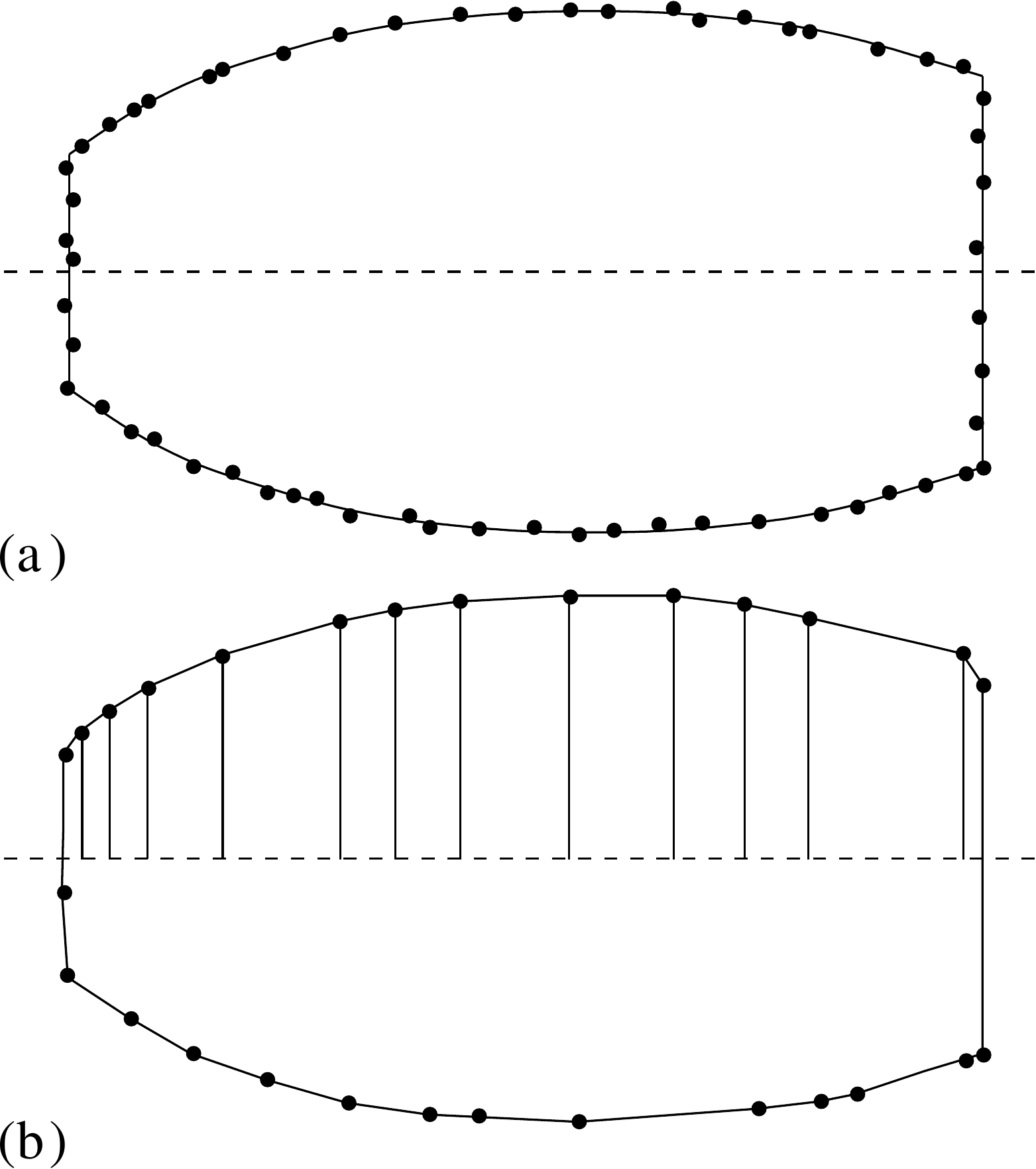}
\caption{\label{fig:cutout}Graph (a) shows an example of the boundary points of the intersection of a colloidal cylinder with a planar interface, which are obtained using the triangular tessellation scheme, together with its convex hull (full curves). In (a) all points are shown, in (b) only those on the convex hull, with which the surface area $S_{12}$ is computed through a trapezoidal integration scheme as shown for the area above the $x$-axis. The convex hull also serves to compute the length of the contact line $L$.}
\end{figure}

From the set of triangles $\tilde{\Delta}_{p}$, the points where the plane intersects the original tessellation are extracted. These points form a two dimensional data set which approximates the surface area that is cut out of the interface by the presence of the colloid, see Fig.~\ref{fig:cutout}, from which $S_{12}$ and $L$ can be computed. 

Typically the boundary consists of several hundred grid points, depending on the size of the triangular mesh. The approximate surface area $\tilde{S}_{12}$ is obtained by means of a trapezoidal integration scheme, which is applied to the points above and below the $x$-axis after sorting them by increasing $x$-coordinate. Here we assume that the colloid is convex. The excluded area $S_{12}$ is always a connected set when the colloid is convex. For non-convex colloids the area excluded from the interface can consist of two or more disjoined pieces, e.g., for a dumbbell. To avoid such difficulties we restrict ourselves to uni-axial convex colloids and refer the reader to Appendix A for a more general algorithm. It should, however, be noted that in the case of a dumbbell the surface areas are parts of circles and spheres and therefore this shape can in principle be handled analytically~\cite{unpub}.

The trapezoidal integration scheme suffers from instabilities due to small numerical uncertainties, which potentially interfere with the sorting algorithm, as illustrated in Fig.~\ref{fig:cutout}a. These problems can easily be overcome by considering the convex hull of the data set, Fig.~\ref{fig:cutout}b, thereby eliminating such ``noise''. Considering the convex hull does come at the price of reducing the number of data points. However, for several hundred grid points this effect is negligible. The convex hull of the boundary points is also used to determine $\tilde{L}$, the approximate length of the contact line $L$.

In summary, we have introduced an explicit tessellation scheme to compute the surface areas $S_{1}$, $S_{2}$, and $S_{12}$ as well as the contact line length $L$ explicitly, for any convex (uni-axial) colloid particle. This scheme can, however, be straightforwardly generalized to more complex colloidal shapes with surface heterogeneities. Non-convex shapes and particles with surface patterns are discussed Appendix A.

\section{\label{sec:resu}Results and Discussion}

\begin{figure}
\includegraphics[height=3.0in]{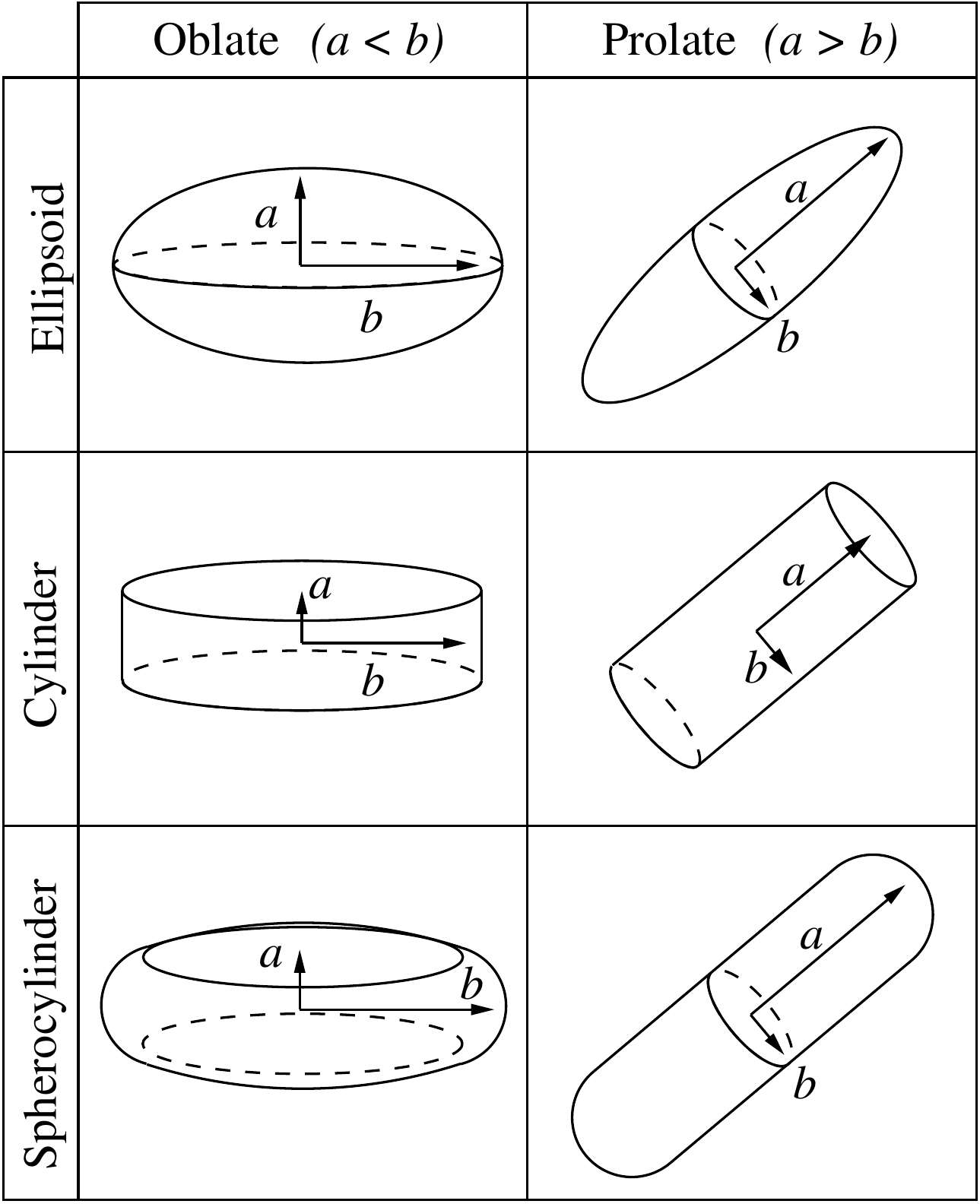}
\caption{\label{fig:objects} Impression of the various colloidal shapes considered in this paper. The left column represents oblate particles ($a < b, m < 1$), the right column prolate particles ($a > b, m > 1$). Note the difference in shape between oblate and prolate spherocylinders.}
\end{figure}

In this section we discuss the adsorption free energy landscape for several particle shapes, namely ellipsoids, cylinders, and spherocylinders, as shown in Fig.~\ref{fig:objects}. We focus on these three types of particle, as they are frequently used to model colloidal platelets and rod-like colloids in theoretical work and computer simulations. In addition, these particle shapes can be either \emph{prolate} $(m > 1)$ or \emph{oblate} $(m \le 1)$, and have relatively simple parameterizations. Note that in our model the spherocylinder has length $a$ and width $b$. Contrary to traditional notation $a$ includes the spherical end-caps for a prolate spherocylinder, while $b$ includes the rounded side for an oblate spherocylinder. 

Only the dimensionless adsorption free energy, Eq.~\cref{eq:pierpot}, is considered and the investigation here limits itself to two aspect ratios $m = 1/4$ and $m = 4$, one contact angle $\cos \theta = -1/2$, and several values of $\tau^{*}$. Negative values for $\cos\theta$ are used, since then the particle prefers $M_{1}$. However, there is no real difference between $\cos\theta$ and $-\cos\theta$, because it only implies interchanging $M_{1}$ and $M_{2}$ in our model. That is to say, when $(z^{*}_{\mathrm{ad}},\phi_{\mathrm{ad}})$ is the location of a minimum for $\cos \theta > 0$, then for $-\cos \theta$ there is a minimum at $(-z^{*}_{\mathrm{ad}},\phi_{\mathrm{ad}})$ if $\tau^{*}$ is fixed. It should be noted that the line tension can be both positive and negative~\cite{t3} and can assume values in a range spanning several decades~\cite{tension,t1,t2} depending on the precise details of the system. The line tension $\tau^{*}$ is therefore chosen arbitrarily, without reference to a specific experimental system. We have limited ourselves to several interesting configurations, to prove the accuracy of our method. 

An equidistant mesh of $200\times200$ to $400\times400$ vertices is employed, which via the various parameterizations is mapped onto a heterogeneous triangular tessellation. For future reference we will denote a $200\times200$-vertex grid as a $200^{2}$-vertex grid, for instance. This yields relative fractional accuracies in the range $10^{-3}$ to $10^{-5}$ in $S$, $S_{1}$, $S_{2}$, $S_{12}$, $L$, $r_{1}$, $r_{12}$ and $l$, depending on the object parameterized. These uncertainties are established using semi-analytic values of the surface areas and contact line length, given in Appendix B. The semi-analytic nature refers to the fact that one dimensional integrals need to be evaluated in order to obtain a value. More than $5000$ non-equidistant grid points are used in these calculations to ensure a relative fractional uncertainty lower than $10^{-6}$ (the magnitude is determined using grid reduction). These semi-analytic results are independent of mesh size and triangular tessellation and can be used to test our method. For higher accuracy, meshes of $1000^{2}$ vertices are employed, although in most cases these results are indistinguishable from the $400^{2}$-vertex mesh results, or the semi-analytic results. We are therefore confident that the numerical scheme is sufficiently stable and can be applied to shapes for which we have not performed analytic verification.

\subsection{\label{sub:ellip}Ellipsoids}

\begin{figure*}
\includegraphics[width=6.25in]{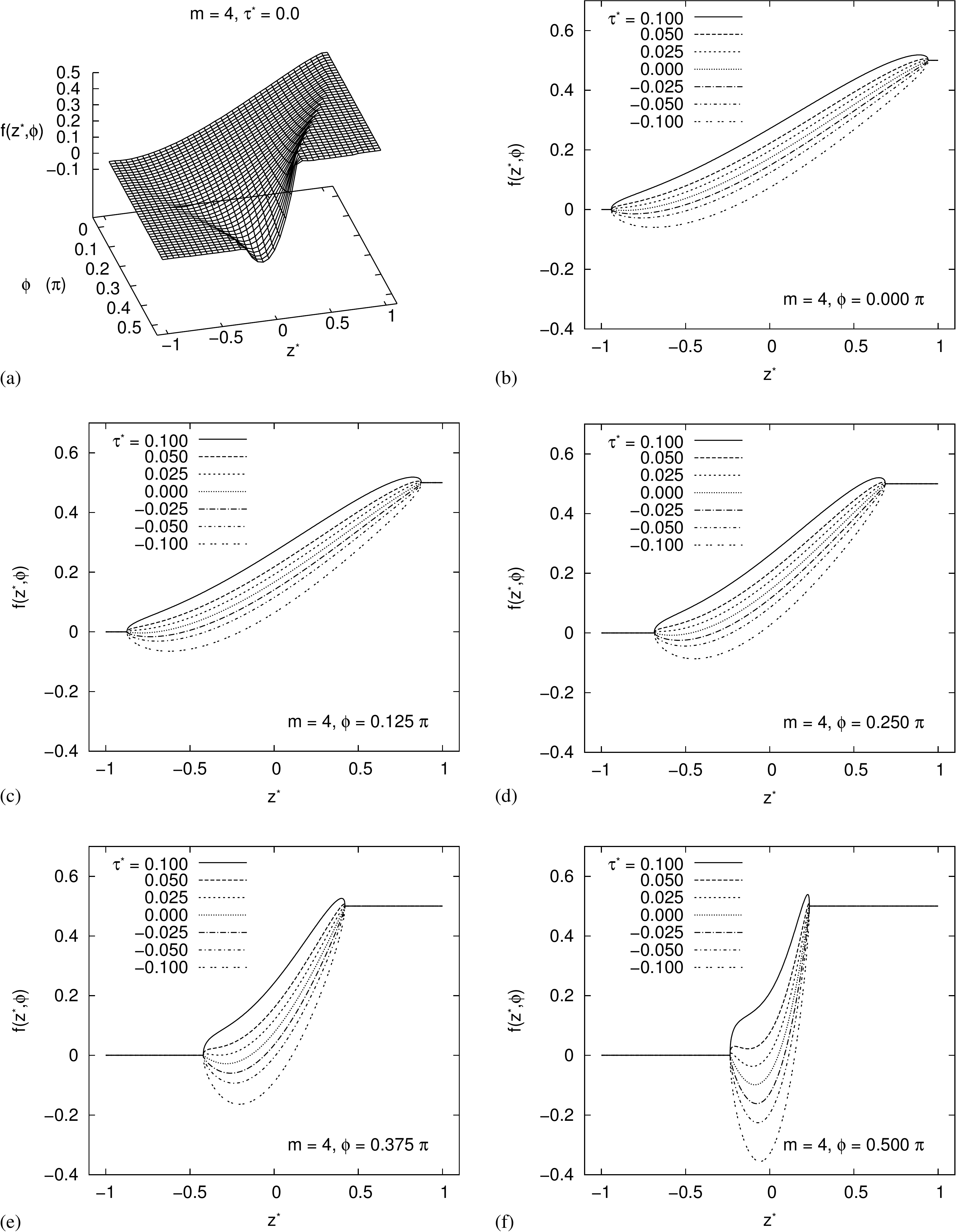}
\caption{\label{fig:ellpot} The adsorption free energy landscape for an ellipsoid with aspect ratio $m = 4$, and contact angle $\cos \theta = -1/2$. Graph (a) shows $f(z^{*},\phi)$ in a 3D representation as a function of $z^{*}$ and $\phi$ for $\tau^{*} = 0$. Graphs (b) through (f) show $\phi$-sections of this free energy landscape for $\phi = 0$, $\pi/8$, $\pi/4$, $3\pi/8$, and $\pi/2$ respectively, and for $\tau^{*} = -0.1$, $-0.05$, $-0.025$, $0$, $0.025$, $0.05$, $0.1$. The central dotted line in each graph corresponds to a section of graph (a), i.e., $\tau^{*} = 0$. The adsorption free energy $\phi$-sections are shown for $-1 < z^{*} < 1$, but can be extended with constant value $0$ for $z \le -1$ and $0.5$ for $z \ge 1$. Note the appearance of free energy barriers for $\tau^{*} > 0$ which must be crossed if the particle adsorbs to the interface. When $\tau^{*} > 0.05$ the adsorption free energy does not have a minimum. In this figure numerical and analytic results agree within the line width of the curves. }
\end{figure*}

We calculate the adsorption free energy $f(z^{*},\phi)$ of an ellipsoid with aspect ratio $m = 4$ and contact angle $\cos \theta = -1/2$. We use a $400^{2}$-vertex mesh for the tessellation and a $5000$-point equidistant trapezium-rule grid for the semi-analytic approach. Figure~\ref{fig:ellpot}a shows $f(z^{*},\phi)$ for $\tau^{*} = 0$ in a 3D representation as a function of $z^{*}$ and $\phi$. In Figs.~\ref{fig:ellpot}(b-f) we plot $\phi$-sections of the free energy landscape for varying $\tau^{*}$. The numerical results established using the tessellation scheme presented above and
 the semi-analytic results obtained using the equations and techniques described in Appendix B agree within the line width of the curves, i.e., the relative uncertainty is lower than $10^{-3}$ for all grid points. Graphs similar to those in Fig.~\ref{fig:ellpot} can be made for ellipsoids, cylinders, and spherocylinders of any aspect ratio, with any contact angle, and line tension. We have verified that the semi-analytic scheme in Appendix B and our tessellation scheme yield the same results.

In the specific case of an ellipsoidal colloid with $m = 4$ and $\cos\theta = -1/2$, see Fig.~\ref{fig:ellpot}, we find that for negative values of $\tau^{*}$ there is a single minimum in all $\phi$-sections of the free energy landscape. From these figures it can also be derived that there is in fact a single minimum in the adsorption free energy for $\tau^{*} < 0$. That is to say, a single minimum at $\phi = \pi/2$ and no meta-stable secondary minima. We will come back to this shortly. From Figs.~\ref{fig:ellpot}(b-f) it is clear that the single minimum in the $\phi$-sections with $-z^{*}_{\mathrm{det}} < z^{*} < z^{*}_{\mathrm{det}}$ for $\tau^{*} < 0$ vanishes with increasing $\tau^{*}$. The minimum free energy for a given $\phi$, is then $f_{\min}(\phi) = 0$ and this minimum is assumed when $z^{*} < -z^{*}_{\mathrm{det}}$. In this case, the particle prefers to be detached from the interface and can move freely in $M_{1}$, where its adsorption free energy is lowest. 

For certain $\tau^{*}$ a $\phi$-section can have two minima, for example $\tau^{*} = 0.05$ and $\phi = \pi/2$, see Fig.~\ref{fig:ellpot}f. Here there is an absolute minimum given by the detached state with $f_{\min}(\pi/2) = 0$ and a meta-stable (local) minimum with $z^{*} \in [-z^{*}_{\mathrm{det}},z^{*}_{\mathrm{det}}]$, i.e., when the particle is adsorbed at the interface. For sufficiently positive $\tau^{*}$ the presence of this local meta-stable minimum in the $\phi$-sections is dependent on the value of $\phi$. There is a local minimum with $z^{*} \in [-z^{*}_{\mathrm{det}},z^{*}_{\mathrm{det}}]$ for $\phi = \pi/2$ and $\phi = 3\pi/8$ when $\tau^{*} = 0.025$ for instance, but this minimum is not present for the $\phi \le \pi/4$ sections, see Figs.~\ref{fig:ellpot}(b-f). Conversely the minimum with $z^{*}_{\min} \in [-z^{*}_{\mathrm{det}},z^{*}_{\mathrm{det}}]$ can be the absolute minimum and the detached state $(z^{*} < - z^{*}_{\mathrm{det}})$ a meta-stable minimum, see for instance the $\phi$-section in Fig.~\ref{fig:ellpot}f with $\tau^{*} = 0.025$.

The appearance of a local meta stable minimum with $z^{*} \in [-z^{*}_{\mathrm{det}},z^{*}_{\mathrm{det}}]$ is in part related to the formation of ``adsorption barriers'' in the free energy. That is to say, when $\tau^{*}$ is sufficiently positive, the colloid has to cross a barrier to attach at the interface from an immersed state in either medium. From Figs.~\ref{fig:ellpot}(b-f) we observe that positive values for the line tension $\tau^{*}$ give rise to these adsorption barriers. This behavior is most clearly visible in the $\tau^{*} = 0.025$ and $\tau^{*} = 0.05$ $(\phi = \pi/2)$-sections in Fig.~\ref{fig:ellpot}f. The height of the barriers varies with the value of the polar angle and they become more pronounced with increasing $\tau^{*}$. These barriers are quite intriguing, since they can prevent a particle form reaching its lowest free energy state, when it is initially introduced in its least favored medium. 

The above results agree with the findings in Ref.~\cite{acicular}. However, there are several differences between our results and those of Ref.~\cite{acicular} as the expressions in Appendix B and in Ref.~\cite{acicular} do not agree completely. Apart from minor typographical errors, there is a problem with the definition of subdomains on which the equations hold, as well as the way in which quantities are made dimensionless. Because of notational differences these problems are not immediately obvious, but when comparing results it is clear that the adsorption free energy barriers induced by the line tension are far less pronounced in our case. We believe that the way in which $\tau$ is made dimensionless in Ref.~\cite{acicular} violates scale invariance, but from the description given this is impossible to determine. Despite the discrepancies with previously established results, we are confident that our results are correct, since we have used two independent methods, which yield the same results within numerical uncertainty.

\begin{figure}[h!]
\includegraphics[width=3.375in]{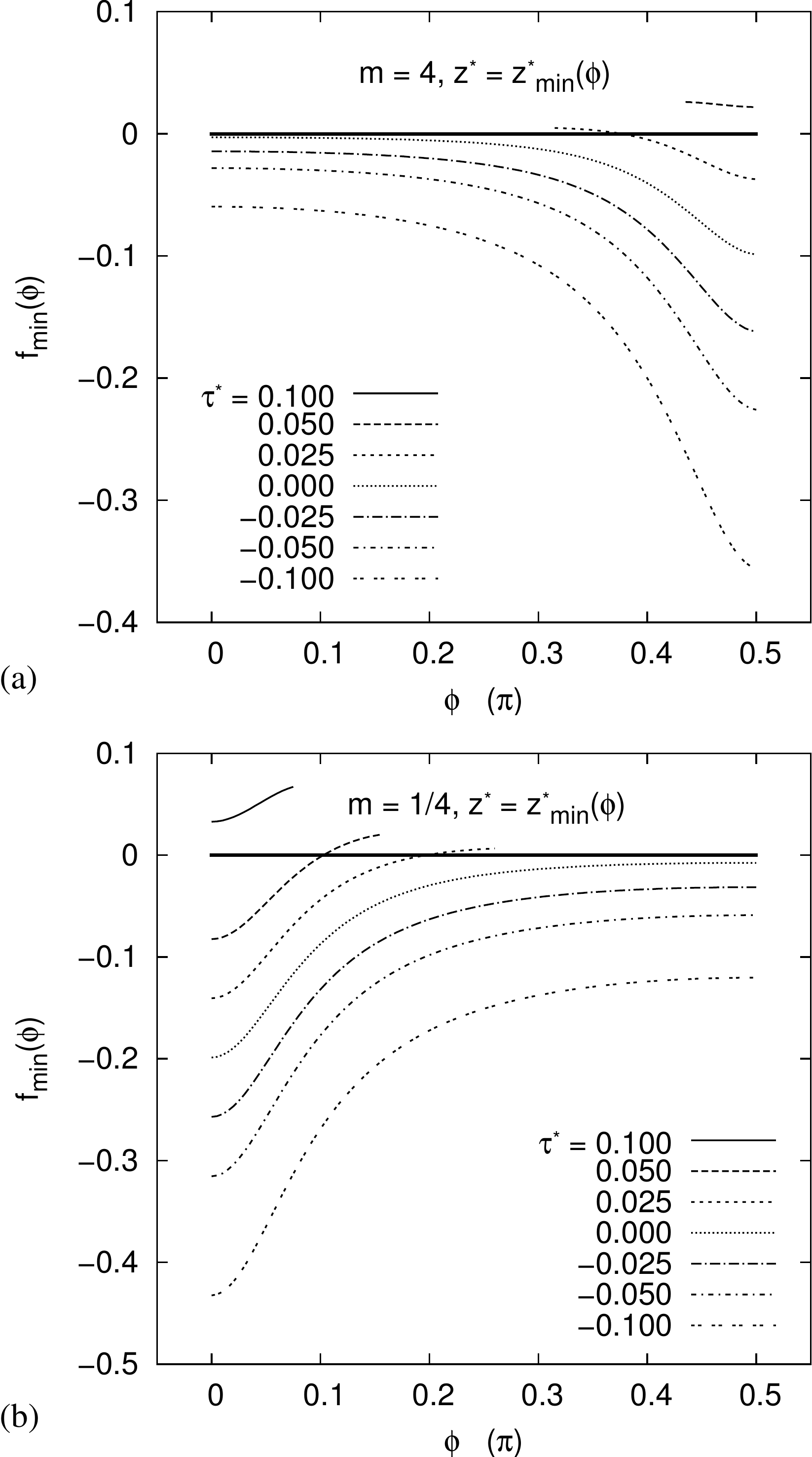}
\caption{\label{fig:ellside} The minimum adsorption free energy $f_{\min}(\phi)$ for ellipsoids with aspect ratio $m = 4$ (a), $m = 1/4$ (b), $\cos \theta = -1/2$, and several values of $\tau^{*}$. The thick line indicates $f_{\min}(\phi) \equiv 0$, which is the minimum free energy when $\tau^{*} \gg 0$. Note that the meta stable part of the adsorption curves is shown as well, i.e., $f_{\min}(\phi) > 0$. We only concern ourselves with the presence of a minimum for a certain $\phi$ when $z^{*} \in [-z^{*}_{\mathrm{det}},z^{*}_{\mathrm{det}}]$, not with its stability. Depending on the value of $\tau^{*}$, a minimum in the $\phi$-sections need not be present along the entire $\phi$-range, also see text. For $\tau^{*}=0.1$ in graph (a) there is no minimum adsorption free energy, in correspondence to the results from Fig.~\ref{fig:ellpot}. Again the agreement between numerical and semi-analytical results is within the line width of the curves.}
\end{figure}

In addition to giving $\phi$-sections, we have determined for which $z^{*}_{\min}$ the adsorption free energy $f(z^{*},\phi)$ is minimal for a given $\phi$. Figure~\ref{fig:ellside} shows the minimum adsorption free energy $f_{\min}(\phi) \equiv f(z^{*}_{\min},\phi)$ for ellipsoids with $m = 4$, $m = 1/4$, $\cos \theta = -1/2$, and several $\tau^{*}$ corresponding to the choices in Fig.~\ref{fig:ellpot}. The minimum adsorption free energy is $f_{\min}(\phi)$ under the constraint that $z_{\min}^{*} \in [-z^{*}_{\mathrm{det}},z^{*}_{\mathrm{det}}]$. The trivial solution $f_{\min}(\phi) = 0$ with $z^{*} < -z^{*}_{\mathrm{det}}$, when $\tau^{*} \gg 0$, is represented by a thick horizontal line. Note that this solution is not in the $[-z^{*}_{\mathrm{det}},z^{*}_{\mathrm{det}}]$ domain. We have also indicated any meta-stable part of the adsorption free energy, i.e., $f_{\min}(\phi) > 0$. These minima are metastable with respect to the $f_{\min}(\phi) = 0$ solution, for which $z^{*} < -z^{*}_{\mathrm{det}}$. This local minimum need not exist for all values of $\phi$, which results in the minimum adsorption free energy curves terminating when $\tau^{*} > 0.025$ $(m = 4)$ and $\tau^{*} > 0.0$ $(m=1/4)$ respectively, see Fig.~\ref{fig:ellside}. This is in agreement with the behavior of the $\phi$-sections given Fig.~\ref{fig:ellpot}(d-f). For $\tau^{*} < 0.025$ all minimum adsorption free energy curves are stable.

If there is a minimum adsorption free energy curve for a given $\tau^{*}$, then we find that the (local) minimum of the free energy is located somewhere on this curve, per definition. For $m = 4$ the location of the the adsorption minimum of $f$ corresponds to $\phi_{\mathrm{ad}} = \pi/2$. Similarly, when $m = 1/4$ the adsorption free energy $f$ is minimal for $\phi_{\mathrm{ad}} = 0$. This observation proves to hold in general for ellipsoids, because the $\phi_{\mathrm{ad}}$ is determined solely by the aspect ratio~\cite{unpub}. We alluded to this result earlier, in describing Fig.~\ref{fig:ellpot}, but the presence of a single minimum and the monotonicity of the minimum adsorption free energy curves is much more evident in Fig.~\ref{fig:ellside}. To summarize, the effect of positive line tension is found to be two fold. Firstly, it gives rise to an adsorption barrier, which must be overcome for particles to attach to the interface. Secondly, it destabilizes the adsorption of a colloid, by reducing the depth of the free energy minimum at adsorption, even to the point that it is either meta-stable or non-existent. The angular dependence of this stability reduction can be attributed to the anisotropy in the particle shape. 

\subsection{\label{sub:cylin}Cylinders}

\begin{figure}
\includegraphics[width=3.375in]{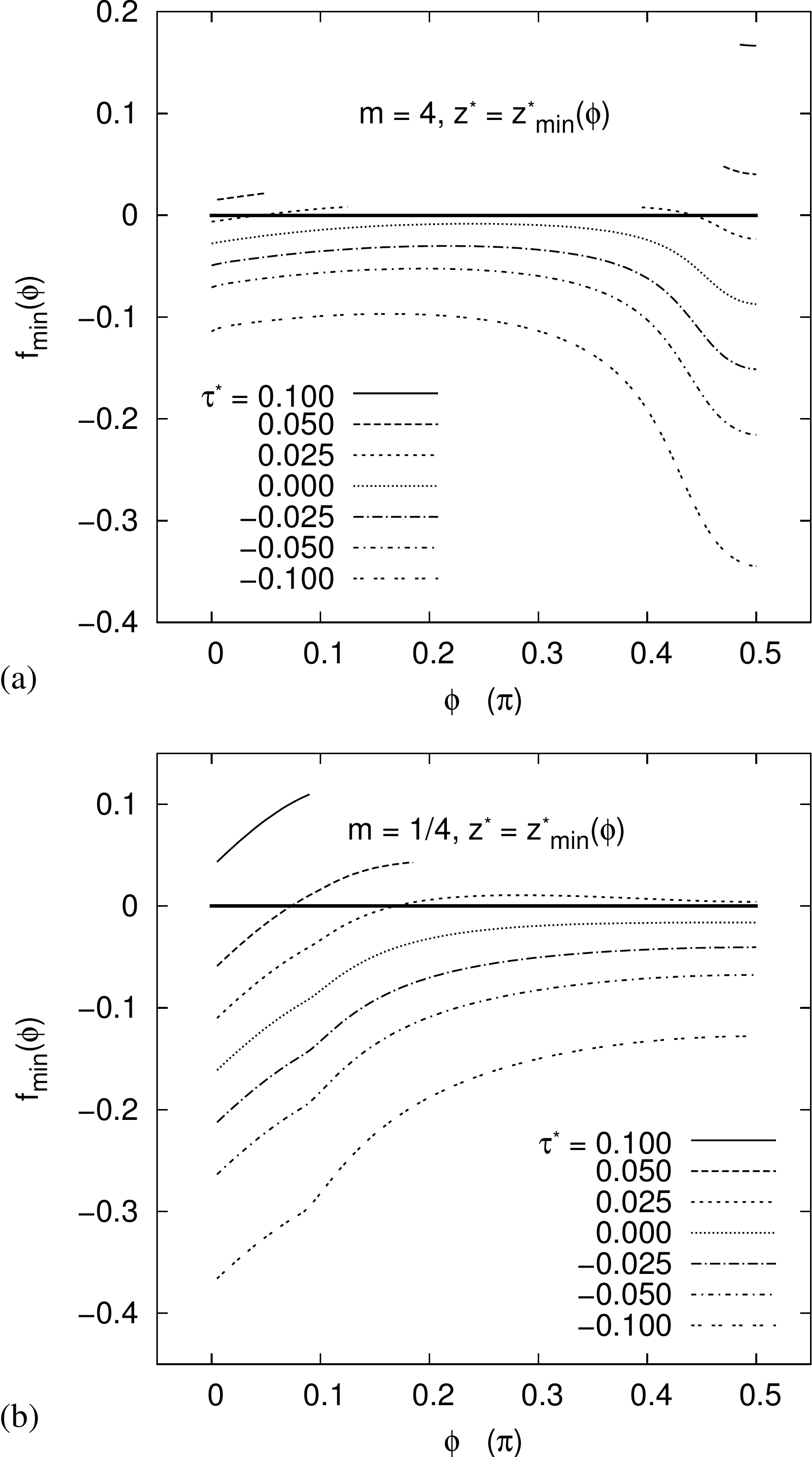}[h]
\caption{\label{fig:cylside} The minimum adsorption free energy curves $f_{\min}(\phi)$ with $\cos \theta = -1/2$ and several values of $\tau^{*}$ for a cylinder with $m = 4$ (a), and $m = 1/4$ (b). The thick line indicates $f_{\min}(\phi) \equiv 0$, which is the stable minimum when $\tau^{*} \gg 0$. The agreement between numerical and semi-analytical results is within the line width of the curves. Note the appearance of a secondary minimum in $f_{\min}(\phi)$. The kink in the minimum free energy curves (graph b) for the oblate cylinder is caused by the sharp edge of the cylinder, see text.}
\end{figure}

We only study the minimum adsorption free energy as a function of the polar angle $\phi$ for cylindrical particles and several $\tau^{*}$ when $\cos\theta = -1/2$. Figure~\ref{fig:cylside} shows the corresponding curves $f_{\min}(\phi)$ for cylinders with $m = 4$ (a) and $m = 1/4$ (b). Again we find that for $\tau^{*} > 0.1$ there is only one minimum in the free energy with $z_{\min}^{*}< -z_{\mathrm{det}}^{*}$, namely when the colloid is completely immersed in $M_{1}$. It can be shown that
 this trend holds in general. When comparing Fig.~\ref{fig:cylside}a with the results given in Ref.~\cite{acicular}, there is no correspondence. This lack in accordance can be attributed to the fact that the equations presented in Ref.~\cite{acicular} do not sample all possible orientations of the cylinder with respect to the interface. Again our results have been verified by comparison with the semi-analytic result. The agreement is better than that the line width of the curves in Fig.~\ref{fig:cylside} can show. 

The cylindrical particles have two noticeable differences with their ellipsoidal counterparts. Firstly, we find that two minima can appear in a single minimum adsorption free energy curve. Whereas for ellipsoids these curves have monotonic properties, and the free energy therefore has only one minimum, the cylindrical curves can have two minima. For the configurations considered the absolute minimum is located at $\phi_{\mathrm{ad}} = \pi/2$ for $m = 4$, i.e., the particle lies flat on the interface. The meta-stable minimum is at $\phi_{\mathrm{ad}} = 0$, i.e., the particle is perpendicular to the interface. The latter corresponds to a cylinder which only has one of its disk shaped end-caps flush with the interface and the rest of its surface in $M_{1}$. The free energy gained by excluding a disk from the interface and not having any other part in contact with $M_{2}$, which is energetically unfavorable, can be sufficient to generate a local meta-stable minimum. It is also possible to choose parameters such that this ``odd'' configuration has the lowest adsorption free energy and is therefore stable. Secondly, we also observe the presence of a kink in some of the minimum adsorption free energy curves, see Fig.~\ref{fig:cylside}b. This feature is related to the sharp corners of the cylinder itself, which gives rise to ridges in the adsorption free energy landscape. For the two configurations considered here, only the minimum free energy curves for $m = 1/4$ follow parts of these ridges, and consequently a kink appears. For the prolate cylinder the minimum is found away from the ridges in the free energy landscape and the kinks do not appear in the minimum adsorption free energy curves.

\subsection{\label{sub:sphecy}Spherocylinders}

\begin{figure}[h!]
\includegraphics[width=3.375in]{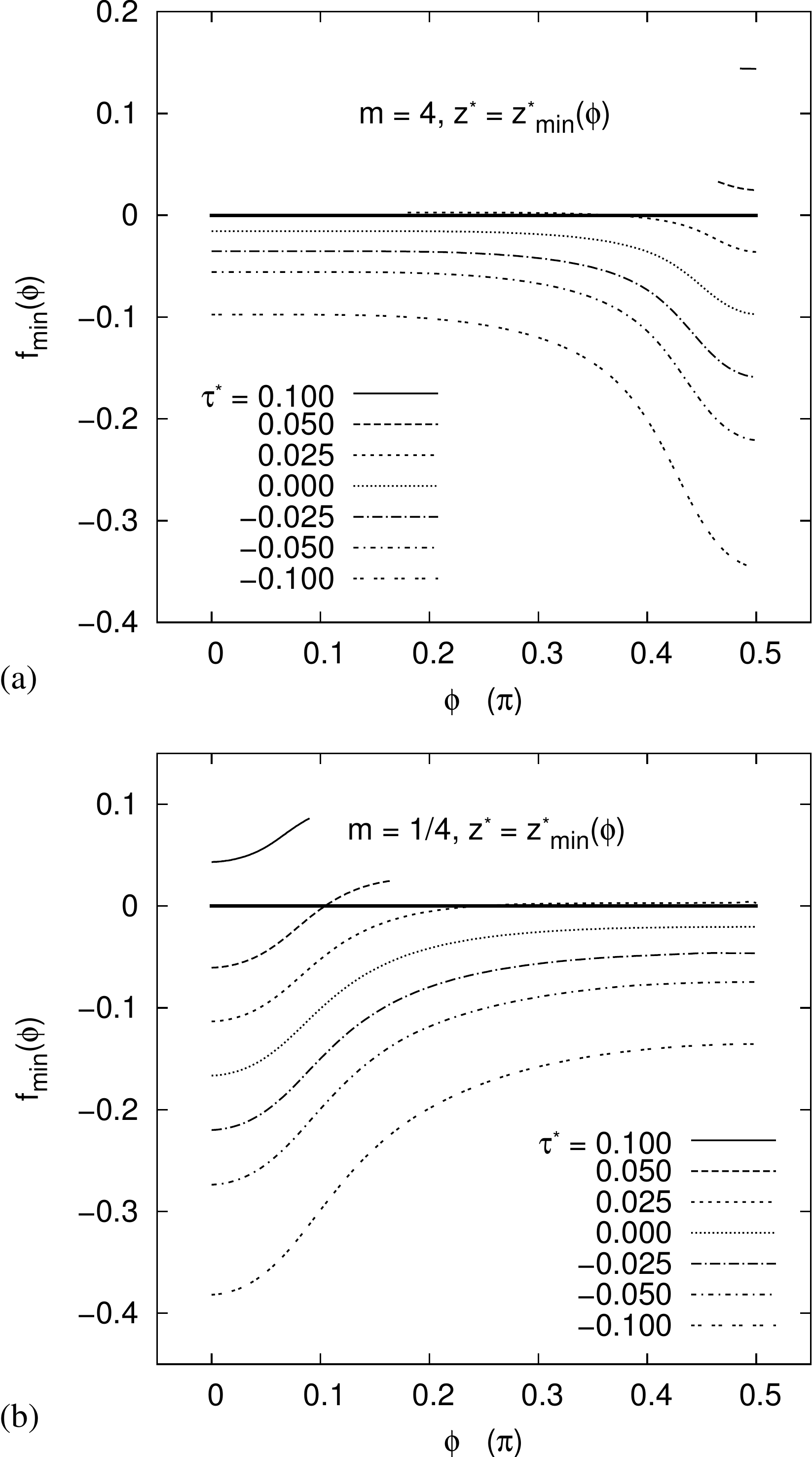}
\caption{\label{fig:spcside} The minimum adsorption free energy curves $f_{\min}(\phi)$ with $\cos \theta = -1/2$ and several values of $\tau^{*}$ for a spherocylinder with $m = 4$ (a), and $m = 1/4$ (b). The notations are analogous to those in Figs.~\ref{fig:ellside} and~\ref{fig:cylside}. Again the agreement between the two methods used to generate the results is better than the line width of the curves. Note that in both graphs there is only one $\phi_{\mathrm{ad}}$ per minimum curve.}
\end{figure}

For spherocylinders we again concentrate on determining for which $z^{*}_{\min}$ the adsorption free energy $f(z^{*},\phi)$ is minimal for a given $\phi$, under the constraint that $z_{\min}^{*} \in [-z^{*}_{\mathrm{det}},z^{*}_{\mathrm{det}}]$. The minimum adsorption free energy curves $f_{\min}(\phi)$ for spherocylindrical particles with $m = 4$ (a), $m = 1/4$ (b), $\cos \theta = -1/2$ are given in Fig.~\ref{fig:spcside} for several values of $\tau^{*}$. Note that for spherocylinders these curves are similar to those for ellipsoids. The agreement between numerical and semi-analytic results is again better than can be appreciated from the line width. For the semi-analytic results a grid of 8000 non-equidistant points was used to evaluate the one dimensional integrals using Aitken's method~\cite{aitken}, also see Appendix B. This yields a relative error of $10^{-7}$ based on grid reduction. The numerical tessellation scheme is based on $1000^{2}$ vertices to obtain a relative accuracy of $10^{-5}$ per point or better, when compared to the semi-analytic curves.

There is only one minimum per minimum adsorption free energy curve, as the particle shapes are smooth and in that sense more similar to ellipsoids than to cylinders. For spherocylinders, $\phi_{\mathrm{ad}}$ is also completely determined by the aspect ratio $m$, i.e., $\phi_{\mathrm{ad}} = 0$ for $m <  1$ and $\phi_{\mathrm{ad}} = \pi/2$ for $m > 1$~\cite{unpub}. This property further distinguishes smooth convex uni-axial particles, e.g., ellipsoids and spherocylinders, from non-smooth particles, e.g., cylinders. However, the mechanism behind adsorption to the interface is quite subtle, depending not only on $m$, $S$, $\cos \theta$, and $\tau^{*}$, but to a large degree on the shape and surface patterning of the colloid. A more detailed study on the effect of particle shape and surface patterning will be the topic of future work~\cite{unpub} and will be presented elsewhere.

\section{\label{sec:conc}Concluding Remarks}

We have presented a numerical framework to determine the adsorption free energy, in terms of surface and line tension contributions, of a non-convex patterned colloid adsorbed at a flat interface. This framework is a natural extension of established theoretical models, e.g., Refs.~\cite{pieran0,oettel0,bresme,acicular}. Because of the complexities which arise when determining the adsorption free energy for complex colloidal shapes, a numerical technique based on triangular tessellation is developed. The accuracy and stability of this method has been extensively verified for convex uni-axial colloids, such as ellipsoids, cylinders, and spherocylinders. This analysis was performed by comparing semi-analytic results with the results produced by our tessellation technique. The expressions used to obtain the semi-analytic results are given in Appendix B, and thereby amend previous results~\cite{acicular}. Exact correspondence is found between the numerical and semi-analytical results, and the established analytic values for specific configurations of colloids adsorbed at a flat interface~\cite{bresme}.

Despite the fact that all of the presented results can be derived semi-analytically, the numerical method presented here has substantial merits. On the one hand, the semi-analytic results can be calculated numerically substantially more quickly than the triangular tessellation results. Only the value of one dimensional integrals needs to be approximated for these. On the other hand, the process of determining analytic expressions can be very labor intensive and may result in equations which are incorrect. Therefore, the numerical scheme can be employed to verify the equations derived by analytic means. However, verification is not the true strength of the numerical technique. Determining an appropriate parametrization and corresponding tessellation is much less involved than obtaining the analytic expressions, and hence the numerical method is more suited to quickly determine the adsorption free energy landscape of a range of different shapes and surface patterns. 

It should also be pointed out that, once implemented, the triangular tessellation scheme will have the same numerical limitations for any system it is applied to. By identifying these limitations, they can be avoided, making the method very robust. Semi-analytic results are different in this respect, since they suffer from a lack of uniformity in the numerical techniques that are needed. The diversity of 1D and 2D integrals, which require numerical evaluation, obtained by studying general systems is limitless. The numerical evaluation of these should be scrutinized on a case by case level, a numerical integrator that works for one integral, is not necessarily suited for another. In our experience, the merits of the triangular tessellations scheme, greatly outweigh those of the semi-analytic approach for generality, stability, and applicability. Hence, our method based on such a triangular tessellation scheme, can be used to examine the physics behind interfacial adsorption of general colloidal particles as well as to make a link between theory and experiments~\cite{unpub}.

\section{\label{sec:ackn}Acknowledgments}

MD acknowledges financial support by the ``Nederlandse Organisatie voor Wetenschappelijk Onderzoek'' (NWO) for a Vici Grant, and both MD and RvR thank Utrecht University for financial support by the High Potential Programme.

\appendix

\section{\label{appsec:extended}Extension of Theory and Numerical Scheme to Non-Convex Patterned Particles.}

\subsection{\label{subapp:theory}Theoretical Model}

\begin{figure}
\includegraphics[height=5.0in]{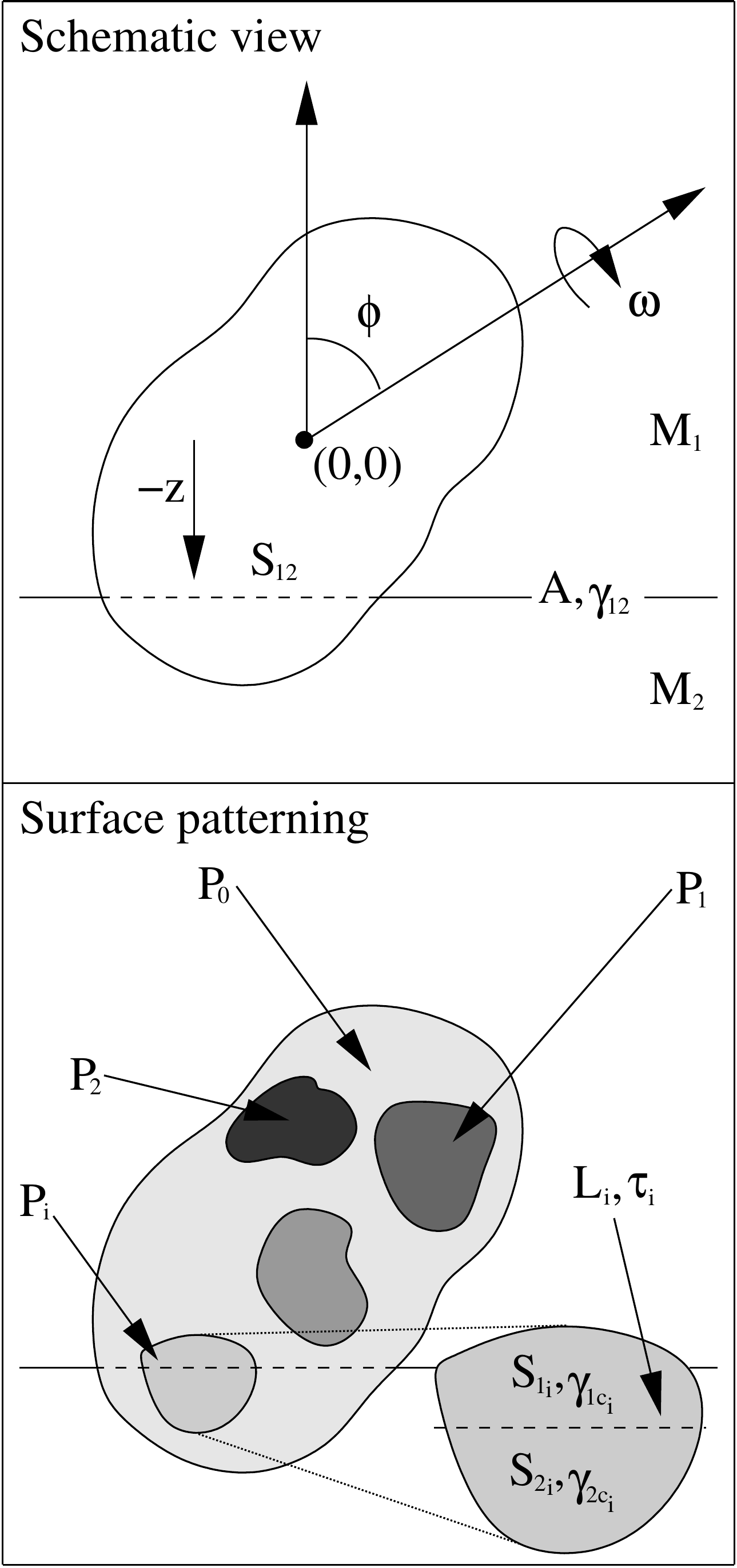}[h]
\caption{\label{fig:patches} A non-convex surface-patterned colloid adsorbed to an interface in schematic representation (top) and the same colloid's surface patterning (bottom). The orientation of the colloid is described by three quantities: (i) the depth $z$; (ii) the angle $\phi$, which some arbitrary but fixed axis through the colloid's center-of-mass makes with the interfacial normal; (iii) the angle $\omega$ which gives rotations around the chosen axis. The angle $\omega$ is defined with respect to a predetermined initial configuration. The origin $(0,0)$ is located at the center-of-mass of the colloid. Again $A$ is the total (macroscopic) surface area of the interface, $S_{12}$ the area excluded from the interface by the presence of the particle, and $\gamma_{12}$ the medium 1 $(M_{1})$ - medium 2 $(M_{2})$ surface tension. The individual patches are labeled by $P_{i}$ (bottom), where $i = 0, \dots n$. Each patch has a $P_{i}-M_{1}$ surface tension $\gamma_{1c_{i}}$, a $P_{i}-M_{2}$ surface tension $\gamma_{2c_{i}}$, and a line tension $\tau_{i}$. The surface area of $P_{i}$ in $M_{1}$ is given by $S_{1_{i}}$, the surface area in $M_{2}$ by $S_{2_{i}}$, and the contact line length by $L_{i}$.}
\end{figure}

In this section we extend our theory to non-convex colloids with surface patterning. We consider a colloid, of which the surface is divided into several areas with different liquid-solid surface tension properties, as is illustrated in Fig.~\ref{fig:patches}. The surface patterning, other than the most basic, breaks the rotational symmetry properties of our system. Hence, an arbitrary non-convex patterned particle is described by an angle $\omega$ in addition to the two parameters $z$ and $\phi$ used earlier, see Fig.~\ref{fig:patches}. Suppose an arbitrary axis is fixed through the center-of-mass of the particle, at which the origin of the system is located, then $\phi$ is the angle between this axis
 and the interfacial normal. The angle $\omega$ describes rotations around this axis. Rotations by $\omega$ are well defined, if they are with respect to some initial orientation. This initial orientation can, however, be arbitrarily chosen. 

Let the particle be partitioned into $n$ \emph{patches} $P_{i}$, with
 the index $i = 0 , 1 , \dots, n$, see Fig.~\ref{fig:patches}. Each patch has a specific patch-medium surface tension $\gamma_{1c_{i}}$ $(P_{i}$-$M_{1})$ and $\gamma_{2c_{i}}$ $(P_{i}$-$M_{2})$, in addition to a line tension $\tau_{i}$. Let $S_{1_{i}}$ be the surface area of patch $P_{i}$ in $M_{1}$, $S_{2_{i}}$ the surface area in $M_{2}$, and $L_{i}$ the length of the contact line. These three quantities can be any value between $0$ and their maximum, depending on the orientation of the colloid. The adsorption free energy of the colloid is given by
\begin{equation}
\label{eq:a_th_shfree} V(z,\phi,\omega) = \gamma_{12}( A - S_{12}) + \sum_{i=0}^{n} \left( \gamma_{1c_{i}} S_{1_{i}} + \gamma_{2c_{i}} S_{2_{i}} + \tau_{i} L_{i} \right). 
\end{equation}
We define $S_{i} \equiv S_{1_{i}} + S_{2_{i}}$, such that the total surface area is $S = \sum_{i = 0}^{n} S_{i}$. As before we can set the adsorption free energy to be zero in $M_{1}$, by subtracting 
\begin{equation}
\label{eq:a_th_shift} \gamma_{12} A + \sum_{i=0}^{n} \gamma_{1c_{i}} S_{i},
\end{equation}
from $V(z,\phi,\omega)$, to obtain the shifted adsorption free energy
\begin{equation}
\label{eq:a_th_patpot} F(z,\phi,\omega) = \sum_{i=0}^{n} \left( (\gamma_{1c_{i}} -  \gamma_{2c_{i}}) (S_{1_{i}} - S_{i}) + \tau_{i} L_{i} \right) - \gamma_{12}S_{12}. 
\end{equation}
The following quantities are introduced analogous to the theoretical description given earlier, $\gamma_{12} \cos \theta_{i} \equiv \gamma_{1c_{i}} -  \gamma_{2c_{i}}$ via Young's equation, $r_{i} \equiv S_{i}/S$, $r_{1_{i}} \equiv S_{1_{i}}/S$, $r_{12} \equiv S_{12}/S$, $l_{i} \equiv L_{i}/\sqrt{S}$, and $\tau_{i}^{*} \equiv \tau_{i} / \gamma_{12} \sqrt{S}$. Using the above definitions Eq.~\cref{eq:a_th_patpot} is reduced to the elegant dimensionless from
\begin{equation}
\label{eq:a_th_poten} f(z^{*},\phi,\omega) = \sum_{i=0}^{n} \left( \cos \theta_{i} (r_{1_{i}} - r_{i}) + \tau^{*}_{i} l_{i} \right) - r_{12},
\end{equation}
where the shifted adsorption free energy has been divided by $\gamma_{12} S$. Here $z$ is made dimensionless by introducing $z^{*} = z/\mathcal{R}$, with $\mathcal{R}$ the radius of the smallest sphere which encloses the particle.

Note that in our model we have glanced over one detail, namely that it is possible to have a four-phase contact line on the boundary of two patches, if the interface coincides (partially) with this boundary. We define the line tension associated to this four-phase contact line to be the average of the line tensions of the two patches, for mathematical convenience. The contribution to the adsorption free energy is the length of this four-phase contact line times the averaged line tension. To the authors' knowledge, little is known about the properties of such four-phase line tensions. There have been studies into four-phase contact lines~\cite{fourphase}, but the line tension is not included. We believe that averaging is not unreasonable. However, multi-phase line tensions and point tension contributions to the adsorption free energy certainly merit further investigation. Sharp features in the colloid, such as cusps and facets, and their possible adsorption free energy contributions due to stresses induced on the interface by surface exclusion are also not considered here.

\subsection{\label{subapp:numeric}Improved Numerical Scheme}

The numerical scheme, based on triangular tessellation as described in the main part of this paper, is suited to handle non-convex shapes, with the exception of the surface area excluded from the interface. In this paragraph we extend the numerical scheme to handle non-convex patterned particles. Note that it is not excluded that these particles have handles. 

First the object is tessellated with triangles using a suitably chosen parametrization. This tessellation obeys the following rules. (i) Patch boundaries are approximated by triangle edges, consequently, a single triangle has a single set of surface properties. (ii) Each triangle is labeled according to the patch it is in, with the label $P_{i}$. (iii) The direction of the surface normal of each triangle is known, and is required to point outwards from the particle. (iv) A sufficiently large number of small triangles is used where the surface of particle changes abruptly, either via a large gradient or a cusp-like feature.

Determining the approximated total surface area $\tilde{S}$ is analogous to the procedure outlined before. Similarly, the surface area of patch $i$, $S_{i}$, is approximated by summing the surface area of triangles with label $P_{i}$, yielding $\tilde{S}_{i}$. The surface area of a patch above the interface $S_{1_{i}}$ is obtained by partitioning the triangle mesh in the manner described earlier. The approximated surface area is denoted $\tilde{S}_{1_{i}}$. The area $\tilde{S}_{2_{i}}$ can also be determined in this way. The equality $S_{i} = S_{1_{i}} + S_{2_{i}}$ also holds in approximated form $\tilde{S}_{i} = \tilde{S}_{1_{i}} + \tilde{S}_{2_{i}}$ and can be used as a consistency check. 

\begin{figure}
\includegraphics[width=3.375in]{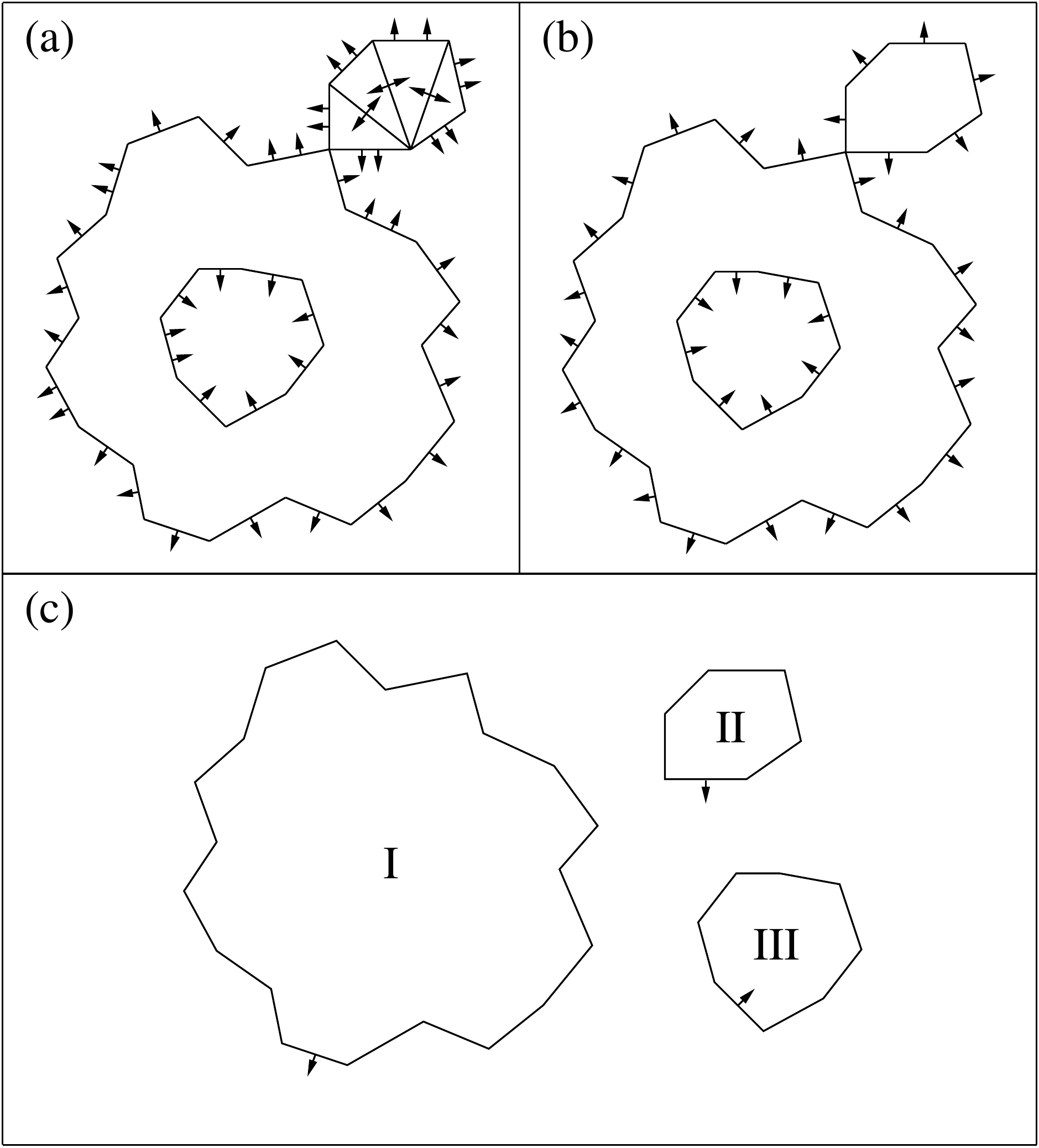}
\caption{\label{fig:loops} Illustration of the loop reduction procedure for the approximated interfacial cut-out $\tilde{S}_{12}$ corresponding to a non-convex colloid. We chose not to complicate the figure by composing $S_{12}$ out of disjoined pieces, but the procedure for this situation is analogous. Box (a) shows the line segments obtained after partitioning triangles, i.e., the set $\Lambda$. Note that some sides have multiple directional unit vectors, indicated by arrows, associated to them. These are double instances of a line segment, which occur whenever triangles have one or more sides flush with the interface as is explained in the text. Box (b) shows $\tilde{\Lambda}$, the set of segments after removing all double and internal line segments. Box (c) shows the subdivision of $\tilde{\Lambda}$ into three closed oriented loops.}
\end{figure}

The calculation of $\tilde{S}_{12}$ and $\tilde{L}_{i}$ is, however, a little more involved. In partitioning the triangles, two points are obtained for each triangle, if the sides of the original triangle are intersected by the interface. These two points span a \emph{line segment}, which is oriented via the normal of the triangle. That is to say, after partitioning a set of line segments, say $\Lambda$, is obtained, see Fig.~\ref{fig:loops}a. The members of $\Lambda$, say $\mathbf{\Lambda}_{i}$ with $i$ an index, are encoded with information on the location of the particle. This encoding is as follows. For a triangle intersected by the interface at an angle, the triangle's normal is projected onto the interface and normalized. This unit vector is referred to as the directional (unit) vector, because it gives orientation to the line segment. A triangle which lies flush with the interface needs to be special cased, here the word flush indicates that all vertices are located in the plane of the interface. The directional unit vectors for each of its sides point outward, the implementation of which is trivial. Each vector $\mathbf{\Lambda}_{i}$ has 7 components, two give the $xy$-location of the starting point of the line segment, two the end point of the segment, two give the direction of the unit vector, and one gives the patch it is associated to. All directional unit vectors obtained in this way, point outward from the colloid (at least locally).

To determine $\tilde{S}_{12}$ the set of line segments $\Lambda$ needs to undergo several refinement steps first. Consider all instances of a $\mathbf{\Lambda}_{i}$, for which there is a $\mathbf{\Lambda}_{j}$, which has the same line segment coordinates, but not necessarily the same directional vector, e.g., see Fig.~\ref{fig:loops}a. The situations in which there are two ``overlapping'' segments are the following. (i) When one of the sides of a triangle is flush with the interface (all points off that side are located in the plane), this gives one line segment. The second segment is given by the triangle which shares that particular side with the original triangle. Both of these segments have the same directional vector. (ii) When an entire triangle is flush with the interface all three sides contribute a line segment. These segments need not necessarily be a part of the boundary of $S_{12}$, or equivalently not a part of the contact line, they may be internal, see Fig.~\ref{fig:loops}a. If one of the sides of the original triangle is an interior side, the adjacent triangle, must also be entirely flush with the interface. Otherwise, it would not be an internal segment, i.e., lie in the interior of $S_{12}$. This ajacent triangle gives a second instance of the line segment, for which the directional vector is opposite to that of the first segment, see Fig.~\ref{fig:loops}a. To eliminate unnecessary segments $\mathbf{\Lambda}_{j}$ is removed from $\Lambda$ if $\mathbf{\Lambda}_{i}$ and $\mathbf{\Lambda}_{j}$ have the same directional vector. If, however, the directional vectors have opposite sign both instances are removed, since then these are interior segments. By subjecting each element in $\Lambda$ to this procedure a new set $\tilde{\Lambda}$ is formed. This set contains only segments which are a part of the boundaries of $S_{12}$, see Fig.~\ref{fig:loops}b. 

The set $\tilde{\Lambda}$ is subdivided into closed \emph{loops}. A loop is determined by choosing a segment in $\tilde{\Lambda}$ and adding its neighbors recursively, until no more new neighbors can be added. This procedure is illustrated as follows. Let $\mathbf{\Lambda}_{i}$ be the starting segment. Then after one iteration we obtain the sequence $\mathbf{\Lambda}_{i - 1} - \mathbf{\Lambda}_{i} - \mathbf{\Lambda}_{i + 1}$, and after $m$ iterations $\mathbf{\Lambda}_{i - m} - \cdots - \mathbf{\Lambda}_{i - 1} - \mathbf{\Lambda}_{i} - \mathbf{\Lambda}_{i
 + 1} - \cdots - \mathbf{\Lambda}_{i + m}$. The last neighbors to be added are either equal, i.e.,
 $\mathbf{\Lambda}_{i-m} = \mathbf{\Lambda}_{i+m}$, in
 which case only one is added, or have a common vertex, in which case the loop is also closed. We thus obtain a loop which is ordered by construction. This procedure is repeated until $\tilde{\Lambda}$ is subdivided into loops. It is a priori not excluded in the above that a loop crosses itself, e.g., a lemniscate like structure. Crossover points are however easily located, by the fact that such a vertex will have attached to it an even number of line segments greater than two. All loops are subsequently cut into closed pieces which do not cross themselves, see Fig.~\ref{fig:loops}c. Let these loops be denoted by $\mathcal{C}_{i}$ with $i$ and index, then $\tilde{\Lambda} = \cup_{\forall i} \mathcal{C}_{i}$ and $\cap_{\forall i} \mathcal{C}_{i} = \emptyset$. 

It is necessary to determine the type of closed loop, i.e., whether it is outward or inward. By construction loops cannot contain both outward and inward segments. For a loop where all the directional vectors point inwards the following holds. Any half-line, starting either end-point of a line segment in the loop, in the direction indicated by the directional vector, will intersect another line segment in the loop. For a loop where all directional vectors point outward, there is at least one line segment for which one of the half-lines drawn through its end-points as before will not intersect another line segment in the loop. Numerically checking this criterion efficiently is not trivial, however, the maximum length over which a half-line needs to be checked for intersection is $2\mathcal{R}$, with $\mathcal{R}$ as in the definition of $z^{*}$. After each loop has been labeled either ``outward'' or ``inward'', the area enclosed by each $\mathcal{C}_{i}$ in $\tilde{\Lambda}$ is easily calculated using a polygonal version of Green's theorem. In practice, it is seldom required to use such a complicated scheme to determine the orientation of the loop. Any knowledge on the possible interfacial cut-outs can be used to make the algorithm more efficient. A dumbbell, for instance, only has outwardly oriented loops, which can be easily derived from its symmetry properties.

The line segments in a loop $\mathcal{C}_{i}$ define a set of 2D points in the plane, which are ordered by the ordering of the loop. Map these points onto three dimensional vectors $\mathbf{a}_{j}$, where the first two components are $xy$-coordinates, the last component is zero, and $j$ is an index. Let the set of these vectors be ordered according to the ordering imposed by $\mathcal{C}_{i}$. If there are $\tilde{n} - 1$ distinct points which define the loop, let $j = 0, \dots , \tilde{n}$ with $\mathbf{a}_{\tilde{n}} = \mathbf{a}_{0}$. The area of $\mathcal{C}_{i}$ is then given by
\begin{equation}
\label{eq:a_num_area} A(\mathcal{C}_{i}) = \left\vert \sum_{j = 0}^{\tilde{n} -1} \mathbf{a}_{j} \times \mathbf{a}_{j+1} \right\vert,
\end{equation}
where the $\times$-symbol indicates the cross product and the vertical bars the norm of the vector obtained by summation. Note that this is indeed an adaptation of Green's integral theorem to polygonal shapes. It can be shown that Eq.~\cref{eq:a_num_area} is only valid when there are no self intersections, which is why these needed to be eliminated first. Define $S(\mathcal{C}_{i})$ to be $+1$ when the orientation of the closed loop is outward and $-1$ if the orientation is inward. The approximated surface area excluded from the interface by the presence of the particle is given by
\begin{equation}
\label{eq:a_num_S12} \tilde{S}_{12} = \sum_{\forall i} S(\mathcal{C}_{i}) A(\mathcal{C}_{i}).
\end{equation}
The calculation of the approximated contact line lengths $\tilde{L}_{i}$ is also possible from the set $\tilde{\Lambda}$, by summing over the lengths of line segments which have the same label $P_{i}$. Note that one needs to special case the instances when the contact line coincides with patch boundaries, as described in the previous section.

In summary, we have presented a method that can be used to determine the free energy of an arbitrary patterned colloid adsorbed to a flat interface. This numerical scheme will be applied in future work~\cite{unpub} on for instance colloidal dumbbells and particles which have Janus-like patterning. There are, however, still open problems, such as point tensions, cusp-effects, and four-phase line tensions which merit further investigation. Finally, it should be noted that in specific cases this scheme can be greatly reduced if properties of the possible interfacial cut-out shapes are known.

\section{\label{appsec:analytic}Analytic Expressions for the Adsorption Free Energy}

In this Appendix we reproduce the analytic expressions for ellipsoids, cylinders, and spherocylinders used to verify the accuracy of the triangular tessellation method. To keep Appendix B concise, derivation of the results is not included. The methods used are however analogous to those applied in \cite{acicular}. Only the expressions for $S$, $S_{1}$, $S_{12}$, and $L$ are considered here. We aim to keep the formulation as general as possible, while at the same time showing the similarities and dissimilarities between the various shapes. 

Some expressions can be reduced in specific cases, i.e., several integrals for ellipsoids can be evaluated to give closed expressions in terms of standard functions. However, such a reduction may result in the expression only holding for oblate and not for prolate particles or vice versa. Some integrals can be reduced using symmetry properties, this will not be done here in order to emphasize similarities. In the case of a spherocylinder the strong difference in shape between oblate and prolate necessitates differentiation between the two aspect ratios. A general equation which describes both types of spherocylinder cannot be given.

It is always implied that the following symmetry properties
\begin{eqnarray}
\label{eq:a_S2} S_{2}(z,\phi) & = & S - S_{1}(z,\phi); \\
\label{eq:a_S1} S_{1}(-z,\phi) & = & S - S_{1}(z,\phi); \\
\label{eq:a_S12} S_{12}(-z,\phi) & = & S_{12}(z,\phi); \\
\label{eq:a_L}	L(-z,\phi) & = & L(z,\phi),
\end{eqnarray}
are used to describe the system and speed up numerical calculation. The equations and variables considered are therefore only given on the $(z,\phi)$-domain
\begin{equation}
\label{eq:a_dom} D = [0,\infty)\times[0,\pi/2].
\end{equation}
It will prove necessary to subdivide this domain into ``disjoined'' pieces on which equations are defined. It can be shown that the equations defined on these subdomains change into each other continuously on the common edges. It can also be shown that these equations reduce to previously established results, Refs.~\cite{bresme,microstr}, when $\phi = 0$ and $\phi = \pi/2$. We do not include these calculations here in the interest of briefness. 

Because of the large number of symbols required to formulate the expressions for $S$, $S_{1}$, $S_{12}$, and $L$, we are forced to recycle notations on a paragraph by paragraph basis. However, an attempt is made to use the same symbols for similar quantities as much as possible. The same holds for the definitions of the subdomains for the various species of particle. To further reduce the notation the dependence of variables on $z$ and $\phi$ is often implicit. In the case that a parameter has a different value on several subdomains, it is implied that any function depending on this parameter should be evaluated with the appropriate value.

All particles considered analytically require the numerical evaluation of one dimensional integrals in order to calculate the various surface areas or the contact line length. Therefore, we refer to this method as semi-analytic. For ellipsoidal, cylindrical, and prolate spherocylindrical particles a simple equidistant trapezoidal scheme can be implemented with relatively small numerical error. However, a more stable technique is required for oblate spherocylinders, due to divergences in some of the integrands near the integration boundary points. A midpoint scheme gives reasonable results, although, we found that an application of Aitken's method, see Ref.~\cite{aitken}, near the boundaries, combined with a trapezoidal scheme in the non-divergent section yields more accurate and stable results.

\subsection{\label{subapp:ellipsoids}Ellipsoids}

In the case of an ellipsoidal particle there are three subdomains which partition $D$ namely
\begin{eqnarray}
\label{eq:a_el_D1a} D_{11} & = & [0,p_{1}] \times [0,\pi/2]; \\
\label{eq:a_el_D1b} D_{12} & = & [p_{1},p_{2}] \times [0,\pi/2]; \\
\label{eq:a_el_D2} D_{2} & = & [p_{2},\infty) \times [0,\pi/2],
\end{eqnarray}
where 
\begin{eqnarray}
\label{eq:a_el_r1} p_{1} & = & a \cos \phi; \\
\label{eq:a_el_r2} p_{2} & = & \sqrt{a^{2}\cos^{2}\phi + b^{2}\sin^{2}\phi}.
\end{eqnarray}
These boundaries give $z_{\mathrm{det}}(\phi)$ and a transition point in the integration domain, where there is a change in integration kernel. It should be noted that the definition of subdomain here is slightly convoluted, since $p_{1}$ and $p_{2}$ depend on $\phi$. The notation $[0,p_{1}] \times [0,\pi/2]$ means that for a specific $\phi \in [0,\pi/2]$ the $z$ domain is the line segment $[0,p_{1}(\phi)]$.

Let us introduce the following parameters, which correspond to two coordinates of the plane-ellipsoid intersection
\begin{eqnarray}
\label{eq:a_el_x} x_{\pm} & = & \frac{-b^{2}z\tan\phi \pm a b \sqrt{p_{2}^{2}-z^{2}}}{(a^{2}+b^{2}\tan^{2}\phi)\cos\phi}; \\[0.5em]
\label{eq:a_el_y} y_{\pm} & = & \frac{a^{2}z \pm a b \tan\phi \sqrt{p_{2}^{2}-z^{2}}}{(a^{2}+b^{2}\tan^{2}\phi)\cos\phi}.
\end{eqnarray}
Using the above definitions, the semi-axes of the ellipsoidal cutout, and $S_{12}$ are determined. The long semi-axis is given by 
\begin{equation}
\label{eq:a_el_acr} a_{\mathrm{cross}} = \frac{1}{2}\sqrt{\left( y_{+} - y_{-} \right)^{2} + \left( x_{+}
 - x_{-}
 \right)^{2}},
\end{equation}
and the short semi-axis by 
\begin{equation}
\label{eq:a_el_bcr} b_{\mathrm{cross}} = b \sqrt{1 - \left( \frac{y_{+} + y_{-}}{2a} \right)^{2} - \left( \frac{x_{+} + x_{-}}{2b} \right)^{2}}.
\end{equation}
Let us further define the integral kernels
\begin{eqnarray}
\label{eq:a_el_ker1} \mathcal{I}_{1}(\eta) & = & 2 \pi a b \sqrt{1 - \left(1 - \left( \frac{b}{a} \right)^{2}
 \right)\eta^{2}}; \\[0.5em]
\label{eq:a_el_ker2} \mathcal{I}_{2}(\eta) & = & \frac{\mathcal{I}_{1}(\eta)}{\pi} \arccos \left( \frac{z - a \eta \cos \phi}{b \sin \phi \sqrt{1 - \eta^{2}}}\right), \quad
\end{eqnarray}
which can be applied to both oblate and prolate particles. Using Eqs.~\cref{eq:a_el_ker1} and~\cref{eq:a_el_ker2}, the following expressions are obtained
\begin{eqnarray}
\label{eq:a_el_I1} \mathcal{J}_{1} & = & \int_{(y_{+}/a)}^{1} \mathcal{I}_{1}(\eta) \ud \eta; \\[0.5em]
\label{eq:a_el_I2} \mathcal{J}_{2} & = & \int_{(y_{-}/a)}^{(y_{+}/a)}  \mathcal{I}_{2}(\eta) \ud \eta,
\end{eqnarray}
which are related to the surface area $S_{1}$, as we will show now.

The total surface area of an ellipsoidal particle is now given by
\begin{equation}
\label{eq:a_el_S} S = \int_{-1}^{1} \mathcal{I}_{1}(\eta) \ud \eta,
\end{equation}
the surface area above the interface by
\begin{equation}
\label{eq:a_el_S1} S_{1}(z,\phi) = \left\{ \begin{array}{cl} 
\mathcal{J}_{1} + \mathcal{J}_{2} &(z,\phi) \in  D_{11} \\
\mathcal{J}_{2} & (z,\phi) \in D_{12} \\ 
0 & (z,\phi) \in D_{2} \\
\end{array} \right. ,
\end{equation}
and the cut-out surface area by
\begin{equation}
\label{eq:a_el_S12} S_{12}(z,\phi) = \left\{ \begin{array}{cl} 
\pi a_{\mathrm{cross}} b_{\mathrm{cross}} &(z,\phi) \in  D_{11} \cup D_{12} \\
0 & (z,\phi) \in D_{2} \\
\end{array} \right. .
\end{equation}
To simplify the equation for the contact line length the following notations are introduced
\begin{eqnarray}
\label{eq:a_el_c} c_{\mathrm{cross}}^{2} & \equiv & \frac{a^{2}_{\mathrm{cross}} - b^{2}_{\mathrm{cross}}}{a^{2}_{\mathrm{cross}}} ; \\[0.5em]
\label{eq:a_el_Lc} L_{\mathrm{cross}} & = & 4 a_{\mathrm{cross}} \int_{0}^{\pi/2} \sqrt{1 - c_{\mathrm{cross}}^{2} \sin^{2}\psi} \ud \psi . \quad
\end{eqnarray}
so that 
\begin{equation}
\label{eq:a_el_L} L(z,\phi) = \left\{ \begin{array}{cl} 
L_{\mathrm{cross}} & (z,\phi) \in  D_{11} \cup D_{12} \\
0 & (z,\phi) \in D_{2} \\
\end{array} \right. .
\end{equation}
Equations~\cref{eq:a_el_c}-\cref{eq:a_el_L} hold for both oblate and prolate ellipsoids, provided $c_{\mathrm{cross}}^{2}$ is allowed to assume negative values, when the particle is oblate.

\subsection{\label{subapp:Cylinders}Cylinders}

For cylinders it is necessary to distinguish between two regimes in polar angle, separated by $\tilde{\phi} = \arctan(a/b)$, both of which have three $z$-domains. The angle $\tilde{\phi}$ gives the natural angle corresponding to the ratio of sides, which determines whether or not the plane can intersect the shaft of the cylinder without intersecting one of the end-caps. Some subdomains can be merged, which leads to the following partitioning of $D$
\begin{eqnarray}
\label{eq:a_cy_D1a} D_{11} & = & [0,p_{1}] \times [0,\tilde{\phi}]; \\
\label{eq:a_cy_D1b} D_{12} & = & [0,p_{1}] \times [\tilde{\phi},\pi/2]; \\
\label{eq:a_cy_D2} D_{2} & = & [p_{1},p_{2}] \times [0,\pi/2]; \\
\label{eq:a_cy_D3} D_{3} & = & [p_{2},\infty) \times [0,\pi/2],
\end{eqnarray}
where 
\begin{eqnarray}
\label{eq:a_cy_r1} p_{1} & = & \left\{ \begin{array}{cl} 
a \cos \phi - b \sin \phi & \phi \in [0,\tilde{\phi}] \\[0.5em]
b \sin \phi - a \cos \phi & \phi \in [\tilde{\phi},\pi/2] \\ 
\end{array} \right. ; \\
\label{eq:a_cy_r2} p_{2} & = & a \cos \phi + b \sin \phi ,
\end{eqnarray}
give the $z$-boundaries. These boundaries represent $z_{\mathrm{det}}(\phi)$ and the position of the edge between the shaft and cap of a cylinder. 
 
The following parameters are introduced to aid notation
\begin{eqnarray}
\label{eq:a_cy_q} q_{\pm} & = & \frac{z \pm a \cos \phi}{b \sin \phi}; \\
\label{eq:a_cy_u} u_{\pm} & = & \arccos q_{\pm} - q_{\pm}\sqrt{1 - q_{\pm}^{2}}; \\
\label{eq:a_cy_v} v_{\pm} & = & q_{\pm} \arccos q_{\pm} - \sqrt{1 - q_{\pm}^{2}}; \\
\label{eq:a_cy_w} w_{\pm} & = & b^{2}( u_{\pm} \pm 2 v_{\pm} \tan \phi  ).
\end{eqnarray}
The parameter $q_{\pm}$ is related to the intersection of the plane and the end-cap, the others are parts of evaluated integrals. The integral kernel
\begin{equation}
\label{eq:a_cy_ker} \mathcal{K}(\psi) =  \frac{2b}{\cos \phi}\sqrt{1 - \sin^{2}\phi \sin^{2} \psi},
\end{equation}
is defined to help determine the contact line length.

Using the above definitions, the following equations for $S$, $S_{1}$, $S_{12}$, and $L$, are derived. The total surface area is
\begin{equation}
\label{eq:a_cy_S} S = 2 \pi b^{2} + 4 \pi a b.
\end{equation}
The area of the colloid's surface above the interface is given by
\begin{equation}
\label{eq:a_cy_S1} S_{1}(z,\phi) = \left\{ \begin{array}{cl}
\pi b^{2}(1 - 2 q_{-} \tan \phi) & (z,\phi) \in D_{11} \\
w_{+} + w_{-} &  (z,\phi) \in D_{12} \\
w_{-} & (z,\phi) \in D_{2} \\ 
0 & (z,\phi) \in D_{3} \\
\end{array} \right. ,
\end{equation}
and the area of the cylinder-plane intersection is given by
\begin{equation}
\label{eq:a_cy_S12} S_{12}(z,\phi) = \left\{ \begin{array}{cl}
\displaystyle \frac{\pi b^{2}}{\cos\phi} & (z,\phi) \in D_{11} \\[1.5em]
\displaystyle \frac{b^{2}(u_{-} - u_{+})}{\cos\phi} &  (z,\phi) \in D_{12} \\[1.5em]
\displaystyle \frac{b^{2}u_{-}}{\cos\phi} & (z,\phi) \in D_{2} \\[1.5em]
0 & (z,\phi) \in D_{3} \\
\end{array} \right. .
\end{equation}
The contact line length is found using the following ``set of equations''
\vspace{25 mm}
\begin{widetext}
\begin{equation}
\label{eq:a_cy_L} L(z,\phi) = \left\{ \begin{array}{cl}
\displaystyle \int_{-\pi/2}^{\pi/2}  \mathcal{K}(\psi) \ud \psi & (z,\phi) \in D_{11} \\[1.5em]
\displaystyle \int_{\arcsin q_{-}}^{\arcsin q_{+}} \mathcal{K}(\psi) \ud \psi + 2b\sqrt{1 - q_{+}^2} + 2b\sqrt{1 - q_{-}^2} &  (z,\phi) \in D_{12} \\[1.5em]
\displaystyle \int_{\arcsin q_{-}}^{\pi/2} \mathcal{K}(\psi) \ud \psi + 2b\sqrt{1 - q_{-}^2} & (z,\phi) \in D_{2} \\[1.5em]
0 & (z,\phi) \in D_{3} \\
\end{array} \right. .
\end{equation}
\end{widetext}
The above equations hold for both oblate and prolate cylindrical particles.

\subsection{\label{subapp:spherocylinders}Spherocylinders}

For spherocylinders the situation is even more complicated than it is for cylinders. Recall that in this paper we deviate from the classical definition of aspect ratio for spherocylinders, by including the caps in the length of a prolate particle and the toroidal rim in the width of an oblate particle. Hence, there are terms proportional to $(a - b)$ present in the equations for prolate spherocylinders, which correspond to the traditionally used length, and $(b - a)$ terms for oblate particles, which correspond to the traditionally used width (also see main text).

Since we did not succeed in formulating a single set of equations which holds for both oblate and prolate spherocylinders, we have split this paragraph into two parts. The first part describes prolate particles, the second oblate particles. For both species there are again two polar angle regimes, each having four $z$-regimes. This makes the notation in the following quite heavy, especially because a large number of parameters is introduced to formulate these equations as elegantly as possible.

\subsubsection{\label{subsubapp:prolate}Prolate}

In the case of a prolate spherocylinder there are two polar
 angle regimes, separated by the angle $\tilde{\phi} = \arctan((a - b)/b)$, each of which can be split into four $z$-domains. Again $\tilde{\phi}$ is the natural transition angle related to the ratio of sides of the cylindrical part of the particle. After reduction $D$ can be written as $D = D_{11} \cup D_{12} \cup D_{2} \cup D_{3} \cup D_{4}$, with
\begin{eqnarray}
\label{eq:a_spp_D1a} D_{11} & = & [0,p_{1}] \times [0,\tilde{\phi}]; \\
\label{eq:a_spp_D1b} D_{12} & = & [0,p_{1}] \times [\tilde{\phi},\pi/2]; \\
\label{eq:a_spp_D2} D_{2} & = & [p_{1},p_{2}] \times [0,\pi/2]; \\
\label{eq:a_spp_D3} D_{3} & = & [p_{2},p_{3}] \times [0,\pi/2]; \\
\label{eq:a_spp_D4} D_{4} & = & [p_{3},\infty) \times [0,\pi/2],
\end{eqnarray}
where
\begin{eqnarray}
\label{eq:a_spp_r1} p_{1} & = & \left\{ \begin{array}{cl} (a - b)\cos \phi - b\sin \phi & \phi \in [0,\tilde{\phi}] \\[0.5em] b \sin \phi - (a - b)\cos \phi & \phi \in [\tilde{\phi},\pi/2] \\ \end{array} \right. ;\quad \quad \\ 
\label{eq:a_spp_r2} p_{2} & = & (a - b)\cos \phi + b\sin \phi; \\
\label{eq:a_spp_r3} p_{3} & = & (a - b)\cos \phi + b . \\
\end{eqnarray}
The three boundary values for $z$ correspond to $z_{\mathrm{det}}(\phi)$, the transition between the shaft and a partially intersected sphere-cap, and the transition between to the situation where only the sphere-cap is intersected and not the shaft.

Let us now redefine some variables. Note that in the following $q_{\pm}$, $u_{\pm}$, and $v_{\pm}$ play a similar role as for the cylinder.
\begin{eqnarray}
\label{eq:a_spp_q} q_{\pm} & = & \frac{z \pm (a - b) \cos \phi}{b \sin \phi}; \\
\label{eq:a_spp_s} s_{\pm} & = & \cos^{2} \phi \sqrt{1 + (1 - q^{2}_{\pm})\tan^{2}\phi} \pm q_{\pm} \sin^{2} \phi; \quad \quad \\
\label{eq:a_spp_t} t_{\pm} & = & \arcsin q_{\pm} + q_{\pm} \sqrt{1 - q_{\pm}^{2}}; \\
\label{eq:a_spp_u} u_{\pm} & = & \arccos q_{\pm} - q_{\pm} \sqrt{1 - q_{\pm}^{2}}; \\
\label{eq:a_spp_v} v_{\pm} & = & q_{\pm} \arccos q_{\pm} - \sqrt{1 - q_{\pm}^{2}}; \\
\label{eq:a_spp_w} w_{\pm} & = & q_{\pm} \cos \phi \sqrt{1 - q_{\pm}^{2}} ; \\
\label{eq:a_spp_x} x_{\pm} & = & \sqrt{1 - q_{\pm}^{2} \sin^{2} \phi}; \\
\label{eq:a_spp_x2} x_{\pm}^{2} & \equiv & 1 - q_{\pm}^{2} \sin^{2} \phi ; \\
\label{eq:a_spp_y} y_{\pm} & = & \arccos \left( \frac{q_{\pm} \cos\phi}{x_{\pm}} \right); \\
\label{eq:a_spp_lp} \lambda_{+} & = & x_{+}y_{+}; \\
\label{eq:a_spp_lm} \lambda_{-} & = & x_{-}(\pi - y_{-}); \\
\label{eq:a_spp_mp} \mu_{+} & = & x_{+}^{2} y_{+}; \\
\label{eq:a_spp_mm} \mu_{-} & = & x_{-}^{2} (\pi - y_{-}),
\end{eqnarray}
where $q_{\pm}$ and $s_{\pm}$ are intersection related quantities, and the other quantities are to simplify the notation of evaluated integrals. The value $x_{\pm}^{2}$ is introduced here, because $x_{\pm}^{2}$ is defined as $1 - q_{\pm}^{2} \sin^{2} \phi$, rather than $\vert 1 - q_{\pm}^{2} \sin^{2} \phi \vert$. The appearance of the absolute value in the latter would be a natural consequence of taking the square of $x_{\pm}$. In this way the notation here is similar to $c_{\mathrm{cross}}^{2}$ for ellipsoids, for which we also used a ``$\equiv$''~symbol. We realize that this notation is somewhat unconventional, but there are instances where $x_{\pm}^{2} < 0$ in our definition. However, it should be pointed out that taking the square root to determine $x_{\pm}$ is not a problem, because on the domains that $x_{\pm}^{2}$ is negative the equations do not contain instances of $x_{\pm}$. The following integral kernels
\begin{eqnarray}
\label{eq:a_spp_ker1} \mathcal{K}(\psi) & = & \frac{2b}{\cos \phi}\sqrt{1 - \sin^{2}\phi \sin^{2} \psi} ; \\[0.5em]
\label{eq:a_spp_ker2} \mathcal{L}_{\pm}(h) & = & 2 b^{2} \arccos \left( \frac{( h \mp q_{\pm})\tan \phi}{\sqrt{1 - h^{2}}} \right) ,
\end{eqnarray}
are (re)defined. Here the $\mp$-sign in Eq.~\cref{eq:a_spp_ker2} is the cause of an asymmetry which appears in the integration boundaries later.

Using the above equations and kernels the following physical quantities are derived, which determine the adsorption free energy
\begin{widetext}
\begin{equation}
\label{eq:a_spp_S} S = 4 \pi a b ;
\end{equation}
\begin{equation}
\label{eq:a_spp_S1} S_{1}(z,\phi) = \left\{ \begin{array}{cl}
\displaystyle 2\pi b^{2}(1 - q_{-}\tan\phi) & (z,\phi) \in D_{11} \\[1.5em]
\displaystyle \int_{-q_{-}}^{s_{-}} \mathcal{L}_{-}(h) \ud h - \int_{q_{+}}^{s_{+}} \mathcal{L}_{+}(h) \ud h + \pi b^{2}(2 - q_{+} - q_{-}) + 2 b^{2}(v_{+} - v_{-}) \tan \phi  & (z,\phi) \in D_{12} \\[1.5em]
\displaystyle \int_{-q_{-}}^{s_{-}} \mathcal{L}_{-}(h) \ud h + \pi b^{2}(1 - q_{-}) - 2 b^{2} v_{-} \tan \phi  & (z,\phi) \in D_{2} \\[1.5em]
\displaystyle 2 \pi b (p_{3} - z) & (z,\phi) \in D_{3} \\[1.5em]
\displaystyle 0 & (z,\phi) \in D_{4} \\
\end{array} \right. ;
\end{equation}
\begin{equation}
\label{eq:a_spp_S12} S_{12}(z,\phi) = \left\{ \begin{array}{cl}
\displaystyle \frac{\pi b^{2}}{\cos(\phi)} & (z,\phi) \in D_{11} \\[1.5em]
\displaystyle \frac{b^{2}(t_{+} - t_{-})}{\cos(\phi)} + b^{2}( \mu_{+} - w_{+} ) + b^{2}(\mu_{-} + w_{-}) & (z,\phi) \in D_{12} \\[1.5em]
\displaystyle  \frac{b^{2} u_{-}}{\cos \phi} + b^{2}(\mu_{-} + w_{-}) & (z,\phi) \in D_{2} \\[1.5em]
\displaystyle \pi b^{2}x_{-}^{2} & (z,\phi) \in D_{3} \\[1.5em]
\displaystyle 0 & (z,\phi) \in D_{4} \\
\end{array} \right. ;
\end{equation}
\begin{equation}
\label{eq:a_spp_L} L(z,\phi) = \left\{ \begin{array}{cl}
\displaystyle \int_{-\pi/2}^{\pi/2} \mathcal{K}(\psi) \ud \psi & (z,\phi) \in D_{11} \\[1.5em]
\displaystyle \int_{\arcsin q_{-}}^{\arcsin q_{+}} \mathcal{K}(\psi) \ud \psi + 2 b(\lambda_{+} + \lambda_{-}) & (z,\phi) \in D_{12} \\[1.5em]
\displaystyle \int_{\arcsin q_{-}}^{\pi/2} \mathcal{K}(\psi) \ud \psi + 2 b \lambda_{-} & (z,\phi) \in D_{2} \\[1.5em]
\displaystyle 2 \pi b x_{-} & (z,\phi) \in D_{3} \\[1.5em]
\displaystyle 0 & (z,\phi) \in D_{4} \\
\end{array} \right. .
\end{equation}
\end{widetext}
Note the sign asymmetry in the integration boundaries of the $\mathcal{L}_{\pm}$ integrals in Eq.~\cref{eq:a_spp_S1} is induced by the $\mp$-sign in Eq.~\cref{eq:a_spp_ker2}. Attempts to rewrite the integral in such a way that the asymmetry is eliminated lead to results which look contrived and are still asymmetric in a certain way. Although the appearance of asymmetries may seem unphysical, we have extensively verified that these equations indeed hold. 

\subsubsection{\label{subsubapp:oblate}Oblate}

For oblate particles none of the integral equations, which describe the surface areas and contact line length, can be evaluated to obtain closed analytic expressions in terms of standard functions. Again 5 relevant domains are found, where the $\phi$-domain is split by $\tilde{\phi} = \arctan(a/(b-a))$. The angle $\tilde{\phi}$ is related to the dimensions of the cylindrical core of the prolate spherocylinder. The subdomains are given by 
\begin{eqnarray}
\label{eq:a_spo_D1a} D_{11} & = & [0,p_{1}] \times [0,\tilde{\phi}]; \\
\label{eq:a_spo_D1b} D_{12} & = & [0,p_{1}] \times [\tilde{\phi},\pi/2]; \\
\label{eq:a_spo_D2} D_{2} & = & [p_{1},p_{2}] \times [0,\pi/2]; \\
\label{eq:a_spo_D3} D_{3} & = & [p_{2},p_{3}] \times [0,\pi/2]; \\
\label{eq:a_spo_D4} D_{4} & = & [p_{3},\infty) \times [0,\pi/2],
\end{eqnarray}
where
\begin{eqnarray}
\label{eq:a_spo_r1} p_{1} & = & \left\{ \begin{array}{cl} a \cos \phi - (b - a)\sin \phi & \phi \in [0,\tilde{\phi}] \\[0.5em] (b - a)\sin \phi - a \cos \phi & \phi \in [\tilde{\phi},\pi/2] \\ \end{array} \right. ; \quad \\
\label{eq:a_spo_r2} p_{2} & = & (b - a)\sin \phi + a \cos \phi; \\
\label{eq:a_spo_r3} p_{3} & = & (b - a)\sin \phi + a . \\
\end{eqnarray}
The three $z$-domain $p_{i}$ indicate $z_{\mathrm{det}}(\phi)$, and the position natural transition points on the particle's surface.

To ease notation the following variables are introduced
\begin{widetext}
\begin{eqnarray}
\label{eq:a_spo_xt} x^{t}_{\pm} & = & \cos \phi( z - (b-a)\sin \phi) \pm \sin \phi \sqrt{ (a + z - (b-a) \sin \phi) (a - z + (b-a) \sin \phi)} ; \\[0.5em]
\label{eq:a_spo_xb} x^{b}_{\pm} & = & \cos \phi( z + (b-a)\sin \phi) \pm \sin \phi \sqrt{ (a + z + (b-a) \sin \phi) (a - z - (b-a) \sin \phi)} ; \\[0.5em]
\label{eq:a_spo_z} \mu_{\pm}^{i} & = & 2 \pi a \left( a - x_{\pm}^{i} + (b-a) \arccos \left( \frac{x_{\pm}^{i}}{a} \right) \right),
\end{eqnarray}
\end{widetext}
where $i$ can be either `$t$' or `$b$'. These stand for `top' and `bottom' respectively, but the latter is in no way related to the rotational radius of the particle, which is also given by $b$. The points $x_{\pm}^{t}$ and $x_{\pm}^{b}$ are locations where the interfaces intersects the spherocylinder in a conveniently chosen coordinate frame. The quantities $\mu^{i}_{\pm}$ originate form the evaluation of integrals. The following useful functions are defined to aid notation
\begin{eqnarray}
\label{eq:a_spo_l} h(x) & = & z \sin \phi - \frac{ x - z \cos \phi}{ \tan \phi } \\[0.5em]
\nonumber k(x) & = & (b - a)^{2} \arccos \left( \frac{h(x)}{b-a} \right)  \\
\label{eq:a_spo_k} & & - h(x) \sqrt{(b-a)^{2} - h^{2}(x)}. \\[0.5em]
\label{eq:a_spo_r} r(x) & = & (b - a) + \sqrt{a^{2} - x^{2}} \\
\label{eq:a_spo_y} w(x) & = & \sqrt{ r^{2}(x) - h^{2}(x) } 
\end{eqnarray}
Here $h$ stands for a height related function, $k$ is a integration reduction function, $r$ is a radial distance function, and $w$ is a width function. In this paragraph a prime denotes a derivative with respect to $x$, e.g., $h'(x) \equiv \partial h(x) / \partial x$. For the oblate spherocylinder three integral kernels are required, namely
\begin{eqnarray}
\label{eq:a_spo_ker1} \mathcal{N}_{1}(x) & = & \frac{2 a r(x)}{\sqrt{a^{2} - x^{2}}}\arccos \left( \frac{h(x)}{r(x)} \right) ; \\[0.5em]
\label{eq:a_spo_ker2} \mathcal{N}_{2}(x) & = & \frac{2 w(x)}{\sin \phi} ; \\[0.5em]
\label{eq:a_spo_ker3} \mathcal{N}_{3}(x) & = & 2 \sqrt{1 + (h'(x))^{2} + (w'(x))^{2}} .
\end{eqnarray}
The first of
 the integral kernels is used in determining the area $S_{1}$, the second to determine $S_{12}$, and the third is used to determine the contact line length.

The above equations and kernels are applied to derive the following equations
\begin{widetext}
\begin{equation}
\label{eq:a_spo_S} S = 2 \pi \left( b^{2} + (\pi - 2) ba - (\pi - 3)a^{2} \right) ;
\end{equation}
\begin{equation}
\label{eq:a_spo_S1} S_{1}(z,\phi) = \left\{ \begin{array}{cl}
\displaystyle \int_{x_{-}^{t}}^{x_{+}^{b}} \mathcal{N}_{1}(x) \ud x + \pi(b -a)^{2} + \mu_{+}^{b} & (z,\phi) \in D_{11} \\[1.5em]
\displaystyle \int_{-a}^{a} \mathcal{N}_{1}(x) \ud x + k(a) + k(-a) & (z,\phi)
 \in D_{12} \\[1.5em]
\displaystyle \int_{x_{-}^{t}}^{a} \mathcal{N}_{1}(x) \ud x + k(a) & (z,\phi) \in D_{2} \\[1.5em]
\displaystyle \int_{x_{-}^{t}}^{x_{+}^{t}} \mathcal{N}_{1}(x) \ud x & (z,\phi) \in D_{3} \\[1.5em]
\displaystyle 0 & (z,\phi)
 \in D_{4} \\
\end{array} \right. ;
\end{equation}
\begin{equation}
\label{eq:a_spo_S12} S_{12}(z,\phi) = \left\{ \begin{array}{cl}
\displaystyle \int_{x_{-}^{t}}^{x_{+}^{b}} \mathcal{N}_{2}(x) \ud x & (z,\phi) \in D_{11} \\[1.5em]
\displaystyle \int_{-a}^{a} \mathcal{N}_{2}(x) \ud x & (z,\phi) \in D_{12} \\[1.5em]
\displaystyle \int_{x_{-}^{t}}^{a} \mathcal{N}_{2}(x) \ud x & (z,\phi) \in D_{2} \\[1.5em]
\displaystyle \int_{x_{-}^{t}}^{x_{+}^{t}} \mathcal{N}_{2}(x) \ud x & (z,\phi) \in D_{3} \\[1.5em]
\displaystyle 0 & (z,\phi) \in D_{4} \\
\end{array} \right. ;
\end{equation}
\begin{equation}
\label{eq:a_spo_L} L(z,\phi) = \left\{ \begin{array}{cl}
\displaystyle \int_{x_{-}^{t}}^{x_{+}^{b}} \mathcal{N}_{3}(x) \ud x & (z,\phi) \in D_{11} \\[1.5em]
\displaystyle \int_{-a}^{a} \mathcal{N}_{3}(x) \ud x + 2 w(a) + 2 w(-a) & (z,\phi) \in D_{12} \\[1.5em]
\displaystyle \int_{x_{-}^{t}}^{a} \mathcal{N}_{3}(x) \ud x + 2 w(a) & (z,\phi) \in D_{2} \\[1.5em]
\displaystyle \int_{x_{-}^{t}}^{x_{+}^{t}} \mathcal{N}_{3}(x) \ud x & (z,\phi) \in D_{3} \\[1.5em]
\displaystyle 0 & (z,\phi) \in D_{4} \\
\end{array} \right. .
\end{equation}
\end{widetext}
Note that although the above equations are more symmetric than those for prolate spherocylinders, there is still a degree of asymmetry in the boundary conditions. 

\bibliographystyle{is-unsrt}
\bibliography{BIBL}

\end{document}